\documentclass[useAMS]{mn2e}
\usepackage{natbib}
\usepackage{url,times,graphicx,amsmath,amsfonts,amssymb,aas_macros,epsfig,epstopdf}
\usepackage{multicol}
\usepackage{comment}
\usepackage{bbding} 
\usepackage[usenames,dvipsnames,svgnames,hyperref]{xcolor}
\usepackage[
    pdftex,
    a4paper=true,
    breaklinks=true,
    bookmarks=true,
    bookmarksopen=false,
    bookmarksopenlevel=2
    bookmarksnumbered=true,
    bookmarkstype=toc,
    colorlinks=true,
    citecolor=blue,
    linkcolor=blue,
    menucolor=green,
    urlcolor=magenta,
]{hyperref}

\newcommand{\beq}{\begin{equation}}
\newcommand{\eeq}{\end{equation}}
\def\beqa{\begin{eqnarray}}
\def\eeqa{\end{eqnarray}}

\include{defs}

\def \nifty{nIFTy}

\def \ahf{{\sc ahf}}
\def \gadget{{\sc Gadget}}

\def \arepo{{\sc AREPO}}
\def \arepoIL{{\sc Arepo-IL}}
\def \arepoSH{{\sc Arepo-SH}}

\def \gx{{\sc G3-X}}

\def \magneticum{{\sc G3-Magneticum}}
\def \pesph{{\sc G3-PESPH}}
\def \music{{\sc G3-Music}}
\def \musicP{{\sc G2-MusicPI}}
\def \owls{{\sc G3-OWLS}}

\def \gxx{{\sc G2-X}}

\def \ramses{{\sc RAMSES}}
\def \RamsesA{{\sc Ramses-AGN}}

\usepackage[T1]{fontenc} 
 \usepackage{aecompl}

\def \ncodes{10}
\def \ngadget{7}

\title[nIFTy Cluster Comparison II]{nIFTy galaxy cluster simulations II: radiative models }
\author[Sembolini et. al]{Federico Sembolini$^{1,2,}$\thanks{E-mail: federico.sembolini@uam.es},
Pascal Jahan Elahi$^{3}$, 
Frazer R. Pearce$^{4}$,
 Chris Power$^{5,6}$, 
\newauthor Alexander Knebe$^{1,2}$,
Scott T. Kay$^{7}$, 
Weiguang Cui$^{5,6}$, 
Gustavo Yepes$^{1,2}$, 
\newauthor Alexander M. Beck,$^{8}$ 
Stefano Borgani,$^{9,10,11}$ 
Daniel Cunnama$^{12}$
Romeel Dav\'e,$^{13,14,15}$  
\newauthor Sean February,$^{16}$ 
Shuiyao Huang$^{17}$
Neal Katz$^{17}$
Ian G. McCarthy,$^{18}$ 
 \newauthor Giuseppe Murante,$^{9}$ 
 Richard D. A. Newton,$^{7}$
 Valentin Perret,$^{19}$ 
Ewald Puchwein,$^{20}$
\newauthor Alexandro Saro,$^{7,21}$ 
Joop Schaye, $^{22}$
Romain Teyssier$^{19}$
\\
$^{1}$Departamento de F\'isica Te\'{o}rica, M\'{o}dulo 8, Facultad de Ciencias, Universidad Aut\'{o}noma de Madrid, 28049 Madrid, Spain\\
$^{2}$Astro-UAM, UAM, Unidad Asociada CSIC\\
$^{3}$Sydney Institute for Astronomy, A28, School of Physics, The University of Sydney, NSW 2006, Australia\\
$^{4}$School of Physics \& Astronomy, University of Nottingham, Nottingham NG7 2RD, UK\\
$^{5}$International Centre for Radio Astronomy Research, University of Western Australia, 35 Stirling Highway, Crawley, Western Australia 6009, Australia\\
$^{6}$ARC Centre of Excellence for All-Sky Astrophysics (CAASTRO)\\
$^{7}$Jodrell Bank Centre for Astrophysics, School of Physics and Astronomy, The University of Manchester, Manchester M13 9PL, UK\\
$^{8}$University Observatory Munich, Scheinerstr. 1, D-81679 Munich, Germany\\
$^{9}$Astronomy Unit, Department of Physics, University of Trieste, via G.B. Tiepolo 11, I-34143 Trieste, Italy\\
$^{10}$INAF - Osservatorio Astronomico di Trieste,  via G.B. Tiepolo 11, I-34143 Trieste, Italy\\
$^{11}$INFN - Sezione di Trieste, via Valerio 2, I-34127 Trieste, Italy\\
$^{12}$Physics Department, University of the Western Cape, Cape Town 7535, Sotuh Africa\\
$^{13}$Physics Department, University of Western Cape, Bellville, Cape Town 7535, South Africa\\
$^{14}$South African Astronomical Observatory, PO Box 9, Observatory, Cape Town 7935, South Africa\\
$^{15}$ African Institute of Mathematical Sciences, Muizenberg, Cape Town 7945, South Africa\\
$^{16}$Center for High Performance Computing, CSIR Campus, 15 Lower Hope Street, Rosebank, Cape Town 7701, South Africa\\
$^{17}$Astronomy Department, University of Massachusetts, Amherst, MA 01003, USA\\
$^{18}$Astrophysics Research Institute, Liverpool John Moores University, 146 Brownlow Hill, Liverpool L3 5RF, UK\\
$^{19}$Centre for Theoretical Astrophysics and Cosmology, Institute for Computational Science, University of Zurich, Winterthurerstrasse 190, 8057 Zurich, Switzerland\\
$^{20}$Institute of Astronomy and Kavli Institute for Cosmology, University of Cambridge, Madingley Road, Cambridge CB3 0HA, UK\\
$^{21}$Excellence Cluster Universe, Boltzmannstr. 2, 85748 Garching, Germany\\
$^{22}$Leiden Observatory, Leiden University, P.O. Box 9513, 2300 RA Leiden, the Netherlands\\
}

\begin{document}
\date{Accepted XXXX . Received XXXX; in original form XXXX}

\pagerange{\pageref{firstpage}--\pageref{lastpage}} \pubyear{2015}
\maketitle
\label{firstpage}
\clearpage

\begin{abstract}
We have simulated the formation of a massive galaxy cluster (M$_{200}^{\rm crit}$ = 1.1$\times$10$^{15}h^{-1}M_{\odot}$) in a $\Lambda$CDM universe using \ncodes\ different codes (\ramses, 2 incarnations of \arepo\ and \ngadget\ of \gadget), modeling hydrodynamics with full radiative subgrid physics. These codes include Smoothed-Particle Hydrodynamics (SPH), spanning traditional and advanced SPH schemes, adaptive mesh and moving mesh codes. Our goal is to study the consistency  between simulated clusters modeled with different radiative physical implementations - such as cooling, star formation and AGN feedback. We compare images of the cluster at $z=0$, global properties such as mass, and radial profiles of various dynamical and thermodynamical quantities. We find that, with respect to non-radiative simulations, dark matter is more centrally concentrated, the extent not simply depending on the presence/absence of AGN feedback. The scatter in global quantities is substantially higher than for non-radiative runs. Intriguingly, adding radiative physics seems to have washed away the marked code-based differences present in the entropy profile seen for non-radiative simulations in \cite{nifty1}: radiative physics + classic SPH can produce entropy cores. Furthermore, the inclusion/absence of AGN feedback is not the dividing line -as in the case of describing the stellar content- for whether a code produces an unrealistic temperature inversion and a falling central entropy profile. However, AGN feedback does strongly affect the overall stellar distribution, limiting the effect of overcooling and reducing sensibly the stellar fraction. \end{abstract}

\noindent

\section{Introduction} \label{sec:introduction}
This paper is a continuation of the {\em nIFTy cluster comparison project} \cite[][]{nifty1}: a study of the latest state-of-the-art hydrodynamical codes using simulated galaxy clusters as a testbed for theories of galaxy formation. Simulations are indispensable tools in the interpretation of astronomical observations of these objects (see for instance \citealt{Borgani11}). Although early $N$-body simulations only modeled the gravitational evolution of collisionless effects of dark matter (\citealt{White76}; \citealt{Fall78}; \citealt{Aarseth79}), these were vital for interpreting galaxy surveys and unveiling the cosmic web of the large scale structure of the Universe. The focus of modern simulations (see e.g. \citealt{Vogelsberger14}; \citealt{Eagle15}) has shifted to modeling galaxy formation in a cosmological context , incorporating the key physical processes that govern galaxy formation and the intra-cluster medium (ICM).  
 
The details of the physical processes that are part and parcel of building a galaxy remain uncertain. Naturally, these processes include the conversion of gas to stars and the feedback of energy and metals from supernovae into the surrounding medium (see e.g. \citealt{Voit05} for a review of the radiative processes which govern the evolution of the baryonic component). Galaxy clusters offer an ideal testbed for the study of 
these processes and their complex interplay, precisely because their enormous size 
encompasses a wide range of relevant scales.
As mentioned before,  the goal of modern simulations is now focused on modeling galaxy formation, incorporating the key physical processes that
drive galaxy formation - such as the cooling of a collisional gaseous 
component (e.g. \citealt{Pearce00}; \citealt{Muanwong01}; \citealt{Dave02}; \citealt{Kay04}; \citealt{Nagai07b}; \citealt{Wiersma09}), the birth of stars 
from cool overdense gas (e.g. \citealt{Springel03}; \citealt{Schaye08}), the growth of black
holes \citep{DiMatteo05}, and the injection of energy into the inter-stellar medium by supernovae (e.g. \citealt{Metzler94}; \citealt{Borgani04}; \citealt{Dave08} ;\citealt{dallavecchia12}) and powerful AGN outflows \citep[e.g.][]{Thacker06a,Sijacki07,Puchwein08,Sijacki08,Booth09,Steinborn15}. These processes span an enormous dynamic range, both spatial and temporal, from the sub-pc scales of black hole growth 
to the accretion of gas on Mpc scales from the cosmic web. %

One of the main issues with radiative simulations of galaxy clusters is that they tend to convert a large fraction of gas into stars. Observationally,  only 10-15 per cent of the baryon component of clusters is expected to be in the stellar phase \citep{Gonzalez07},  but radiative runs which only include stellar feedback are affected by overcooling and usually convert too large a fraction of the gas (above 30 per cent) inside the cluster virial radius into stars \citep{Borgani11}.
\par
Recent work on hydrodyamic simulations has identified AGN feedback as a suitable candidate for overcoming this problem \citep[e.g.][]{Puchwein08,Puchwein10,Fabjan10,McCarthy10,Battaglia12,Martizzi12,Planelles13,Pike14,lebrun14}.
Heating from AGN occurs via the release of energy during accretion of the ICM gas onto a supermassive black hole hosted by the central cluster galaxy: this energy is sufficiently high to remove gas from the inner regions of clusters. At the same time,  AGN heating may also be able to explain the lack of gas in  the central region of dynamically relaxed clusters (the ``cool core'' clusters). Pre-ejection of gas by AGN in the high-redshift progenitors of present-day clusters may also be crucial \citep{McCarthy11}.

This is where the nIFTy cluster comparison project comes in, building on a long history of important comparison studies of simulated clusters (e.g. the Santa Barbara project, \citealt{Frenk99}, hereafter SB99) as well as galaxies (e.g. the Aquila project - \citealt{Scannapieco12} - and the AGORA project - \citealt{Kim14}). All codes and subgrid modules attempt to model the key processes of galaxy formation. In our first paper, \cite{nifty1} (hereafter S15), we addressed a well known issue, first highlighted in SB99: mesh-based and traditional SPH codes produced galaxy cluster entropy profiles that were not in agreement. Grid based codes displayed a constant entropy core whereas traditional SPH codes produces profiles that continued to fall all the way towards the centre. The latter behavior was due to the artificial surface tension and the associated lack of multiphase fluid mixing in classic SPH \cite[e.g.][]{2008MNRAS.387..427W,2009MNRAS.395..180M}. Modern SPH codes attempted to address the lack of mixing through a variety of means: artificial conduction \citep{Price2008,Valda12} and
pressure-entropy formulations \citep{2001MNRAS.323..743R,saitoh13,Hopkins13}. In S15, we clearly showed that modern SPH is able to create clusters with flat entropy cores that are indistinguishable from those generated by mesh-based codes. 

\par 
Here we tackle the subgrid physics implemented in a variety of state-of-the-art codes. We extend the analysis presented in S15 by performing simulations of the same cluster with full physics runs where codes have radiative mechanisms describing gas cooling, star formation, supernova feedback, black hole accretion and AGN feedback. We used \ncodes\ different codes (\ramses, 2 incarnations of \arepo, \ngadget\ of \gadget), allowing each method to choose their favorite radiative processes modeled by subgrid physics. This allows us to study how the different mechanisms, especially star formation and AGN feedback, influence the properties of simulated clusters. We examine the overall cluster environment and we focus our analysis on revisiting the gas entropy profiles.

\par 
The rest of this paper is organized as follows: in Section \ref{sec:codes} we briefly describe the codes used and the subgrid physics adopted by each code along with a brief description of the data set. We then discuss our results in Sections \ref{sec:bulk}-\ref{sec:baryons}: starting with an overview of the bulk properties of the cluster and the effect of radiative physics (Section \ref{sec:bulk}); followed by the dark matter distribution (Section \ref{sec:DM}); we continue our analysis by studying the baryon distribution (Section \ref{sec:baryons}): in Section \ref{sec:gas} we describe key properties of the gas component such as the temperature, entropy and gas fraction, concluding our analysis by presenting the code-to-code differences in the distribution of stars (Section \ref{sec:stars}). We report our conclusions in Section \ref{sec:summary}.

\section{The Codes} \label{sec:codes}
The initial \nifty\ cluster comparison project, as presented in \citet{nifty1}, included 13 codes -- \ramses,  ART, \arepo, {\sc{Hydra}} and 9 variants of the \gadget\ code. In this study, we consider the subset of these codes in which full radiative subgrid physics has been included. A comprehensive summary of the approach taken to solve the hydrodynamic equations in each of these codes can be found in S15; here we provide a brief recap of this summary, with a focus on a description of the sub-grid physics implemented in each code. Table \ref{tab:codes} lists the codes included in this work and their basic characteristics (the definition of modern and classic SPH codes, as well as that of grid-based and moving-mesh codes, is provided in Section 2 of S15).

\begin{table*}
  \caption{List of all the simulation codes participating in the second part of the nIFTy cluster comparison project, feedback models included, stellar (CSF) and AGN, and different versions if present.}
\label{tab:codes}
\begin{center}
\begin{tabular}{ccccll}
\hline
\hline
Type & Code name				& CSF    & AGN & Versions & Reference\\
\hline
\hline
Grid-based				& \ramses\  & Y & Y &\RamsesA\  & \cite{Teyssier11}\\
\hline
Moving mesh & \arepo\ & Y & Y & \arepoIL\ & \cite{Vogelsberger13,Vogelsberger14} \\
 & & Y & N &\arepoSH\  &\\
\hline
Modern SPH  &\gx\ & Y & Y & \\
& \pesph & Y & N & & Huang et al. (in prep.) \\
 & \magneticum & Y & Y & &\cite{Magn14}\\
 \hline
 Classic SPH & \music\ & Y & N  & \music\ & \cite{Sembolini2013}\\
 & & & &\musicP\ & \cite{Piontek11}\\
 &\owls\ & Y & Y & &\cite{Schaye10} \\
 &\gxx\ & Y & Y &  &\cite{Pike14}\\
\hline
\end{tabular}
\end{center}
\end{table*}

\subsection{Mesh-based Codes}
\subsection*{Grid-based}
\paragraph*{{\sc\bf Ramses} (Teyssier, Perret)}
Ramses is an adaptive mesh refinement code.
For fluid dynamics a directionally unsplit, second order Godunov scheme with the HLLC Riemann solver is used.
The N-body solver is an adaptive particle mesh code, for which the Poisson equation is solved using the multi-grid technique.
The grid is adaptively refined on a cell-by-cell basis, following a quasi-Lagrangian refinement strategy
whereby a cell is refined into 8 smaller new cells if its dark matter or baryonic mass grows by more than a factor of eight.
Time integration is performed using an adaptive, level-by-level, time stepping strategy.
Parallel computing is based on the MPI library, with a domain decomposition set by the Peano-Hilbert space filling curve. 

\medskip

\noindent \emph{Cooling \& Heating:} Gas cooling and heating is performed assuming coronal equilibrium with a modification of the \cite{Haardt96} UV background
and a self-shielding recipe based on \cite{Aubert10}, with an exponential cut-off of the radiation flux with critical density $n_{\rm crit}=0.01$~H/cm$^{-3}$. All Hydrogen and Helium cooling and heating processes are included following \cite{Katz96}.
Metal cooling is added using the \cite{Sutherland93} metal-only cooling function at solar metallicity,
multiplied by the local metallicity of the gas in solar units. In this particular project, we use also a temperature floor $T_{*}=10^4$~K to prevent spurious fragmentation of our relatively poorly resolved galactic discs.

\smallskip

\noindent \emph{Star Formation:} Star formation is implemented as a stochastic process using a local Schmidt law, as in \cite{Rasera06}.
The density threshold for star formation was set to $n_{*}=0.1$~H/cc, and the local star formation efficiency per gas free fall time
was set to 5 per cent.

\smallskip
  
\noindent \emph{Stellar Population Properties \& Chemistry:} Each star particle is treated as a single stellar population (SSP) with a \cite{Chabrier03} IMF. Mass and metal return to the gas phase by core collapse supernovae only. A single average metal specie is followed during this process and advected in the gas as a passive scalar, to be used as an indicator of the gas metallicity in the cooling function.
\smallskip

\noindent \emph{Stellar Feedback:} In this project, no feedback processes related to the stellar population are used.

\smallskip

\noindent \emph{SMBH Growth \& AGN Feedback:} 
The formation of SMBH particle is allowed using the sink particle technique
described in \cite{Teyssier11}. When the gas density is larger than the star formation density threshold,
a boost in the Bondi accretion rate is allowed, using the boost function $\alpha = (n/n_{*})^3$ proposed by \cite{Booth09}.
The SMBH accretion rate is never allowed to exceed the instantaneous Eddington limit. SMBH particles are evolved using a direct gravity solver,
to obtain a more accurate treatment of their orbital evolution. SMBH particles more massive than $10^8~M_{\odot}$ are allowed to merge if their relative velocity is smaller than their pairwise scale velocity. Less massive SMBH particles, on the other hand, are merged as soon as they fall within 4 cells from another SMBH particle.
The AGN feedback used is a simple thermal energy dump with $0.1 c^2$ of specific energy, multiplied by the instantaneous SMBH accretion rate.
 .

\subsection*{Moving mesh}
\paragraph*{{\sc\bf Arepo} (Puchwein)}
\noindent Here we use two different versions of \arepo: one including AGN feedback (\arepoIL) and one not
including it (\arepoSH).

\arepo\ uses a Godunov scheme on an unstructured moving Voronoi mesh; mesh cells move
(roughly) with the fluid. The main difference between \arepo\ and traditional Eulerian
AMR codes is that \arepo\ is almost Lagrangian and Galilean invariant
by construction. The main difference between \arepo\ and SPH codes (see next subsection)
is that the hydrodynamic equations are solved with a finite-volume Godunov scheme. The
version of \arepo\ used in this study conserves total energy in the Godunov scheme, rather
than the entropy-energy formalism described in \citet{Springel10}. Detailed descriptions of
the galaxy formation models implemented in \arepo\ can be found in \cite{Vogelsberger13}
and \citet{Vogelsberger14}, but the key features can be summarized as follows (hereafter we describe the features
of \arepoIL, the radiative models used for \arepoSH\ are the same than \music, and are listed later in this section).

\medskip

\noindent \emph{Cooling \& Heating:} Gas cooling takes the metal abundance into
account.  The  metal  cooling  rate  is  computed  for  solar  composition gas and scaled to the total metallicity of the cell. 
Photoionization and photoheating are followed based on the homogeneous UV
background model of \cite{Faucher09} and the self-shielding prescription of \cite{Rahmati13}. In addition to the
homogeneous UV background, the ionizing UV emission of nearby AGN is taken into account.

\smallskip

\noindent \emph{Star Formation:} The formation of stars is followed with a multi-
phase model of the interstellar medium which is based on \cite{Springel03} (hereafter SH03) 
but includes a modified effective equation of state above the star formation threshold, i.e. above
a hydrogen number density of 0.13 cm$^{-3}$

\smallskip
  
\noindent \emph{Stellar Population Properties \& Chemistry:} Each star particle is
treated as a single stellar population (SSP) with a \cite{Chabrier03}
IMF. Mass and metal return to the gas phase by AGB stars, core
collapse supernovae and Type Ia supernovae is taken into account.
Nine elements are followed during this process (H, He, C, N, O, Ne, Mg, Si, Fe).

\smallskip

\noindent \emph{Stellar Feedback:} Feedback by core collapse supernovae is implicitly invoked by the multiphase star formation model. In addition, we
include a kinetic wind model in which the wind velocity scales with the local dark  matter velocity dispersion ($v_{\rm w} \sim$ 3.7$\sigma_{\rm DM,1D}$)
The  mass-loading  is  determined  by  the  available  energy  which
is  assumed  to  be 1.09$\times$10$^{51}$erg per  core  collapse  supernova.
Wind particles are decoupled from the hydrodynamics until they fall below a specific density threshold or exceed a maximum travel
time. This ensures that they can escape form the dense interstellar medium.

\smallskip

\noindent \emph{SMBH Growth \& AGN Feedback:} SMBHs are treated as collisionless sink particles. Particles with a mass of 10$^5h^{-1}$M$_{\odot}$
are seeded into halos once they exceed a mass of 5$\times$10$^{10}h^{-1}$M$_{\odot}$. The BHs subsequently grow by
Bondi-Hoyle accretion with a boost factor of $\alpha$ = 100. The Eddington
limit on the accretion rate is enforced in addition. AGN are
assumed to be in the quasar mode for accretion rates larger than
5 per cent of the Eddington rate. In this case 1 per cent of the accreted rest mass
energy is thermally injected into nearby gas. For accretion rates
smaller than 5 per cent of the Eddington rate, AGN are in the radio mode
in which 7 per cent of the accreted rest mass energy is thermally injected
into spherical bubbles (similar to \citealt{Sijacki07}). Full details
of the black hole model are given in \cite{Sijacki14}.
\medskip

\subsection{SPH Codes}
\subsection*{Modern SPH}

\paragraph*{{\sc\bf Gadget3-X} (Murante, Borgani, Beck)} \gx\ code is a development of the non-public \gadget3\ code.
 It includes an improved SPH scheme,
described in \cite{Beck15}.  Main changes with respect to the
standard \gadget3\ hydro are: (i) an artificial conduction term that
largely improves the SPH capability of following gas-dynamical
instabilities and mixing processes; (ii) a higher-order kernel
(Wendland C4) to better describe discontinuities and reduce
clumpiness instability; (iii) a time-dependent artificial viscosity
term to minimize viscosity away from shock regions. Both pure
hydrodynamical and hydro/gravitational tests on the performance of our
improved SPH are presented in \cite{Beck15}.

\medskip

\noindent \emph{Cooling \& Heating:} Gas cooling is computed for an optically thin gas and takes
into account the contribution of metals, using the procedure of \citet{Wiersma09},
while a uniform UV background is included following the procedure of \citet{Haardt01}.

\smallskip

\noindent \emph{Star Formation:} Star formation is implemented as in \citet{Tornatore07}
, and follows the star formation algorithm of
SH03 -- gas particles above a given density
threshold are treated as multi-phase. The effective model of SH03 describes a self-regulated,
equilibrium inter-stellar medium and provides a star formation rate that depends upon the gas
density only, given the model parameters.

\smallskip
  
\noindent \emph{Stellar Population Properties \& Chemistry:} Each star particle is considered
to be a single stellar population (SSP). We follow the evolution of each SSP, assuming the 
\citet{Chabrier03} IMF. We account for metals produced in the SNeIa, SNeII and AGB phases, and
follow 15 chemical species. Star particles are stochastically spawned from parent gas
particles as in SH03, and get their chemical composition of their parent gas. Stellar
lifetimes are from \citet{Padovani93}; metal yields from \citet{Woosley95} for SNeII,
\citet{Thielemann03} for SNeIa, and \citet{VdHoek97} for AGB stars. 

\smallskip

\noindent \emph{Stellar Feedback:} SNeII release energy into their surroundings, but
this only sets the hot gas phase temperature and, as a consequence, the average SPH
temperature of gas particles. Supernova feedback is therefore modeled as kinetic and
the prescription of SH03 is followed (i.e. energy-driven scheme with a fixed wind velocity of
350 km/s, wind particles decoupled from surrounding gas for a period
of 30 Myr or until ambient gas density drops below 0.5 times the multiphase density threshold).

\smallskip

\noindent \emph{SMBH Growth \& AGN Feedback:}
AGN feedback, follows the model described in \cite{Steinborn15}. 
Nevertheless, while this model includes a Bondi-Hoyle like gas accretion (Eddington
limited) onto SMBH, distinguishing the cold and the hot component
(their Eq. 19), here we {\it only} consider the cold accretion,
using a fudge-factor $\alpha_{cold}=100$ in the Bondi-Hoyle
formula. In other words, $\alpha_{hot}=0$. The radiative efficiency is
variable, and it is evaluated using the model of \citep{Churazov05}.
Such a model outputs separately the AGN mechanical and
radiative power as a function of the SMBH mass and the accretion rate;
however, here we sum up these powers and give the resulting energy to
the surrounding gas, in form of purely thermal energy. We set the
efficiency of AGN feedback/gas coupling to $\epsilon_{fb}=0.05$.

We tuned the parameters of our new hydro scheme using the tests
presented in \cite{Beck15}, and those of the AGN model for
reproducing observational scaling relations between SMBH mass and
stellar mass of the host galaxies. We note that we did not make any
attempt to tune parameters to reproduce any of the observational
properties of the ICM. First results on the application of this code to simulations of galaxy clusters, including the reproduction of the Cool Core/ Non-Cool Core dichotomy, 
can be found in \cite{Rasia15}.

 A black hole (BH) of mass 5$\times$10$^6h^{-1}M_{\odot}$ is seeded at
the centre of each friends-of-friends (FoF) group whose mass exceeds 
2.5$\times$10$^11h^{-1}M_{\odot}$ and which does not already contain a BH.

\medskip

\paragraph*{{\sc\bf Gadget3-PESPH} (February, Dav\'{e}, Katz, Huang)}
This version of \gadget\ uses the pressure-entropy SPH formulation of \citet[][]{Hopkins13}
with a 128 neighbour HOCTS(n=5) kernel and the time-dependent artificial viscosity scheme of
\cite{morris1997}.

\medskip

\noindent \emph{Cooling \& Heating:} Radiative cooling using primordial abundances
is modeled as described in \citet{Katz96}, with additional cooling from metal
lines assuming photo-ionization equilibrium following \citet{Wiersma09}. A
\citet{Haardt01} uniform ionizing UV background is assumed.

\smallskip

\noindent \emph{Star Formation:} Star formation follows the approach set out in
SH03, where a gas particle above a density threshold
of $n_{\rm H} = 0.13$ cm$^{-3}$ is modeled as a fraction of cold clouds embedded in
a warm ionized medium, following \citet{McKee77}. The star formation
rate obeys the \citet{Schmidt59} law and is proportional to $n_{\rm H}^{1.5}$, with
the star formation timescale scaled to match the $z$=0 \citet{Kennicutt98}
relation. In addition, the heuristic model of \citet{Rafieferantsoa14}, tuned to reproduce
the exponential truncation of the stellar mass function, is used to quench star
formation in massive galaxies. A quenching probability $P_Q$, which depends on
the velocity dispersion of the galaxy, determines whether or not star formation
is stopped in a given galaxy; if it is stopped, each gas particle eligible for
star formation first has its quenching probability assessed, and if it is selected
for quenching then it is heated to 50 times the galaxy  virial temperature, which
unbinds it from the galaxy.

\smallskip
  
\noindent \emph{Stellar Population Properties \& Chemistry:} Each star particle
is treated as a single stellar population with a \citet{Chabrier03} IMF throughout.
Metal enrichment from SNeIa, SNeII and AGB stars are tracked, while 4 elements -- C, O,
Si and Fe -- are also tracked individually, as described by \citet{Oppenheimer08}.

\smallskip

\noindent \emph{Stellar Feedback:}  Supernova feedback is assumed to drive
galactic outflows, which are implemented using a Monte Carlo approach analogous
to that used in the star formation prescription. Outflows are directly tied
to the star formation rate, using the relation $\dot{M}_{wind} = \eta \times$SFR,
where $\eta$ is the outflow mass loading factor. The probability for a
gas particle to spawn a star particle is calculated from the subgrid
model described above, and the probability to be launched in a wind is
$\eta$ times the star formation probability. If the particle is
selected to be launched, it is given a velocity boost of $v_w$ in the
direction of $v\times a$, where $v$ and $a$ are the particle
instantaneous velocity and acceleration, respectively.
This is a highly constrained heuristic model for galactic outflows, described
in detail in \citet{Dave13}, which utilizes outflows scalings expected for
momentum-driven winds in sizable galaxies ($\sigma > 75 $km s$^{−1}$), and energy-driven
scalings in dwarf galaxies. In particular, the mass loading factor (i.e. the mass outflow
rate in units of the star formation rate) is $\eta = 150 \text{km s}^{-1} /\sigma$ for
galaxies with velocity dispersion $\sigma > 75 $km s$^{−1}$ , and $\eta = 150 \text{km s}^{-1}
/\sigma^2$ for $\sigma < 75 $km s$^{−1}$.

\smallskip

\noindent \emph{SMBH Growth \& AGN Feedback:} These processes are not included.

\medskip

\paragraph*{{\sc\bf Gadget3-Magneticum} (Saro)}\magneticum\ is an advanced version of \gadget3. In this version,
a higher order kernel based on the bias-corrected, sixth-order
Wendland kernel \citep{Dehnen2012} with 295 neighbors is
included. The code also incorporates a low viscosity scheme to track
turbulence as original described in \cite{Dolag2005} with
improvements following \cite{Beck15}. Gradients are computed
with high-order scheme \citep{Price2012} and thermal conduction is modeled
isotropically at 1/20th of the Spitzer rate \citep{Dolag2004}. The
simulation is run with a time-step limiting particle wake-up algorithm
\citep{pakmor12}. The models adopted for cooling, star formation and stellar feedback are the same that in \gx, but with 
different parameters.
 
\smallskip

\noindent \emph{Cooling \& Heating:} The simulation
allows for radiative cooling according to \citep{Wiersma09} and heating from a uniform time-dependent ultraviolet background
\citep{Haardt01}.  The contributions to cooling from each one of
11 elements (H, He, C, N, O, Ne, Mg, Si, S, Ca, Fe) have been
pre-computed using the publicly available CLOUDY photoionization code
\citep{Ferland98} for an optically thin gas in (photo-)ionization
equilibrium.  
 
\smallskip

\noindent \emph{Star Formation:} We model the interstellar
medium (ISM) by using a subresolution model for the multiphase ISM of
\cite{Springel03}. In this model, the ISM is treated as a
two-phase medium, in which clouds of cold gas form by cooling of hot
gas, and are embedded in the hot gas phase assuming pressure
equilibrium whenever gas particles are above a given threshold
density.  
 
\smallskip

\noindent \emph{Stellar Population Properties \& Chemistry:} We
include a detailed model of chemical evolution according to \cite{Tornatore07}. Metals are produced by SNII, by supernovae type Ia
(SNIa) and by intermediate and low-mass stars in the asymptotic giant
branch (AGB). Metals and energy are released by stars of different
masses, by properly accounting for mass-dependent life-times (with a
lifetime function according to \citealt{Padovani93}), the
metallicity-dependent stellar yields by \cite{Woosley95} for
SNII, the yields by \cite{VdHoek97} for AGB stars,
and the yields by \cite{Thielemann03} for SNIa. Stars of
different masses are initially distributed according to a Chabrier
initial mass function (IMF; \citealt{Chabrier03}).  
 
\smallskip

\noindent \emph{Stellar Feedback:} The hot gas within the multiphase model describing the ISM
is heated by supernovae and can evaporate the cold clouds. A certain
fraction of massive stars (10 per cent) is assumed to explode as
supernovae type II (SNII). The released energy by SNII (10$^{51}$ erg)
triggers galactic winds with a mass loading rate proportional to the
star formation rate (SFR) with a resulting wind velocity of $v_{\rm w}$ = 350 km/s.  

\smallskip

\noindent \emph{SMBH Growth \& AGN Feedback:} Our simulations
include prescriptions for the growth of black holes and the feedback
from active galactic nuclei (AGN) based on the model of \cite{Springel05b} and \cite{DiMatteo05} with the same modifications as
in \cite{Fabjan10} and some new, minor changes as described
below.  The accretion onto black holes and the associated feedback
adopts a sub-resolution model. Black holes can grow in mass by either
accreting gas from their environments, or merging with other black
holes. The gas accretion rate is estimated by the Bondi-Hoyle
Lyttleton approximation, (\citealt{Hoyle1939}; \citealt{Bondi1944};
\citealt{Bondi1952}). The black hole accretion is always limited to the
Eddington rate and a characteristic boost factor of 100 is applied as
only the accretion to large scale is captured.  Unlike in \cite{Springel05b},
 in which a selected gas particle contributes to accretion
with all its mass, we include the possibility for a gas particle to
accrete only with a fraction (1/4) of its original mass. A fraction
r = 0.1 of the accreted mass is converted into energy, and a fraction
f = 0.1 of this energy is then thermally coupled with gas within the
smoothing length of the BH, weighted using the same SPH kernel used
for the hydrodynamics. Following \cite{Sijacki07}, when the
accretion rate drops below a given threshold, it is assumed that there
is a transition from a ÒquasarÓ mode to a ÒradioÓ mode of AGN
feedback, and the feedback efficiency is enhanced by a factor of 4. In
contrast to \cite{Springel05b}, we modify the mass growth of the
BH by taking into account the feedback, e.g., $\Delta M_{BH} \propto
(1-r)$. Other more technical modifications on the BH dynamics with
respect to the original implementation have been included. We refer
the reader to \cite{Dolag15} \& \cite{Magn14} for more
details, where we also demonstrate that the bulk properties of the AGN
population within the simulation are quite similar to the observed AGN
properties.

\medskip
 \subsection*{Classic SPH}
 \medskip

\paragraph*{{\sc\bf Gadget3-OWLS} (McCarthy, Schaye)} This is a heavily modified version of \gadget3\,
using a classic entropy-conserving SPH formulation with a 40 neighbor M3 kernel. 

\medskip

\noindent \emph{Cooling \& Heating:} Radiative cooling rates for the gas are computed on an
element-by-element basis by interpolating within pre-computed tables (generated with the
{\small{CLOUDY}} code; cf. \citealt{Ferland13}) that contain cooling rates as a function of
density, temperature and redshift calculated in the presence of the cosmic microwave
background and photoionization from a \citet{Haardt01} ionizing UV/X-ray
background \citep[further details in ][]{Wiersma09}.

\smallskip

\noindent \emph{Star Formation:} Star formation follows the prescription of
\citet{Schaye08} -- gas with densities exceeding the critical density for the
onset of the thermogravitational instability is expected to be multiphase and to
form stars \citep{Schaye04}. Because the simulations lack both the physics and
numerical resolution to model the cold interstellar gas phase, an effective
equation of state (EOS) is imposed with pressure $P\propto\rho^{4/3}$ for densities
$n_{\rm H}>n_{\ast}$ where $n_{\ast}=0.1{\rm cm}^{-3}$. As described in \citet{Schaye08},
gas on the effective EOS is allowed to form stars at a pressure-dependent rate that
reproduces the observed Kennicutt-Schmidt law \citep{Kennicutt98} by
construction.

\smallskip
  
\noindent \emph{Stellar Population Properties \& Chemistry:} The ejection of metals
by massive- (SNeII and stellar winds) and intermediate-mass stars (SNeIa, AGB stars)
is included following the prescription of \citet{Wiersma09b}. A set of 11 individual
elements are followed (H, He, C, Ca, N, O, Ne, Mg, S, Si and Fe), which represent all
the important species for computing radiative cooling rates.

\smallskip

\noindent \emph{Stellar Feedback:} Feedback is modeled as a kinetic wind
\citep{dallavecchia08} with a wind velocity $v_{\rm w}=600\rm km\,s^{-1}$ and a
mass loading $\eta=2$, which corresponds to using approximately 40 per cent of
the total energy available from SNe for the adopted \citet{Chabrier03} IMF.
This choice of parameters results in a good match to the peak of the
SFR history of the universe \citep{Schaye10}.

\smallskip
 
\noindent \emph{SMBH Growth \& AGN Feedback:} Each black hole can grow either via mergers with
other black holes within the softening length or via Eddington-limited gas accretion, the
rate of which is calculated using the Bondi-Hoyle formula with a modified efficiency,
setting $\beta=2$ as in \citet{Booth09}. The black hole is forced to sit on the local
potential minimum, to suppress spurious gravitational scattering \citep{Springel05n}. Feedback is done by
storing up the accretion energy (assuming $\epsilon_r=0.1$, $\epsilon_f=0.15$) until at least one
particle can be heated to a fixed temperature of $T_{\rm AGN}=10^8 \rm K$ \citep{Booth09}.
A friends-of-friends algorithm is run on the fly and FOF haloes with at least 100 dark matter particles 
(and that do not yet have a black hole particle) are seeded with a black hole particle.
The initial mass of this particle is set to 10$^{-3}$ times the (initial) gas mass.
\medskip

\paragraph*{{\sc\bf Gadget2-X} (Kay, Newton)} This is a modified version of the original
\gadget2\ Tree-PM code that uses the classic entropy-conserving SPH formulation with
a 40 neighbor M3 kernel. A detailed description of the code can be found in
\cite{Pike14}, but its key features can be summarized as follows.

\medskip

\noindent \emph{Cooling \& Heating:} Cooling follows the prescription of
\citet{Thomas92} -- a gas particle is assumed to radiate isochorically
over the duration of its timestep. Collisional ionization equilibrium is assumed
and the cooling functions of \citet{Sutherland93} are used, with the metallicity
$Z$=0 to ignore the increase in cooling rate due to heavy elements. Photo-heating rates
are not included but the gas is heated to a minimum $T=10^4\rm K$ at $z<10$ and
$n_{\rm H}<0.1 {\rm cm}^{-3}$.

\smallskip

\noindent \emph{Star Formation:} Star formation follows the method of \citet{Schaye08};
it assumes an equation of state for the gas with $n_{\rm H}>0.1\,{\rm cm}^{-3}$, with an effective
adiabatic index of $\gamma_{\rm eff}=4/3$ for constant Jeans mass. Gas is converted to
stars at a rate given by the Kennicutt-Schmidt relation \citep[][]{Schmidt59,Kennicutt98},
assuming a disc mass fraction $f_g=1$. The conversion is
done stochastically on a particle-by-particle basis so the gas and star particles have the
same mass.

\smallskip
  
\noindent \emph{Stellar Population Properties \& Chemistry:} Each star particle is
assumed to be a single stellar population with a \citet{Salpeter55} IMF.

\smallskip

\noindent \emph{Stellar Feedback:} A prompt thermal Type II SNe feedback model is used.
This assumes that a fixed number, $N_{\rm SN}$, of gas particles are heated to a fixed
temperature, $T_{\rm SN}$, with values of $N_{\rm SN}=3$ and $T_{\rm SN}$=10$^7K$ chosen to match observed
hot gas and star fractions \citep[cf.][]{Pike14}. Heated gas is allowed to
interact hydrodynamically with its surroundings and radiate.

\smallskip

\noindent \emph{SMBH Growth \& AGN Feedback:} A variation on the \citet{Booth09} model
is used. Black holes are seeded in friends-of-friends (FOF) haloes with more than 50
particles at $z=5$, at the position of the most bound star or gas particle, which is
replaced with a black hole particle. The gravitational mass of the replaced particle is
unchanged but an \emph{internal} mass of $10^6 h^{-1} {\rm M}_{\odot}$ is adopted for the
calculation of feedback. Each black hole can grow either via mergers with other black holes
within the softening length or via Eddington-limited gas accretion, the
rate of which is calculated using the Bondi-Hoyle formula with a modified efficiency,
setting $\beta=2$ as in \citet{Booth09}. The black hole is forced to sit on the local
potential minimum, to suppress spurious gravitational scattering. Feedback is done by
storing up the accretion energy (assuming $\epsilon_r=0.1$, $\epsilon_f=0.15$) until at least one
particle can be heated to a fixed temperature of $T_{\rm AGN}=3 \times 10^8 \rm K$. 
This high temperature was chosen for high-mass clusters to match their observed pressure
profiles -- a lower temperature causes too much gas to accumulate in cluster cores because
there is insufficient entropy to escape to larger radius).

\paragraph*{{\sc\bf Gadget3-MUSIC} (Yepes, Sembolini)} This is the original code adopted for MUSIC-2 dataset \citep{Sembolini2013}, simulated using a  modified version of the \gadget3\
Tree-PM code that uses classic entropy-conserving SPH formulation with a 40 neighbor
M3 kernel. The basic SH03 model was used, the key features of
which can be summarized as follows. In this work we also present {\sc\bf Gadget2-MUSIC}, an alternative version of MUSIC performed using the radiative feedbacks
described in \cite{Piontek11} (\musicP\ since now on).

\medskip

\noindent \emph{Cooling \& Heating:} Radiative cooling is assumed for a gas of primordial
composition, with no metallicity dependence, and the effects of a background homogeneous
UV ionizing field is assumed, following \citet{Haardt01}.

\smallskip

\noindent \emph{Star Formation:} The SH03 model is implemented.

\smallskip
  
\noindent \emph{Stellar Population Properties \& Chemistry:} A \citet{Salpeter55} IMF is
assumed, with a slope of -1.35 and upper and lower mass limits of $40 \rm M_{\odot}$ and
$0.1 \rm M_{\odot}$ respectively.

\smallskip

\noindent \emph{Stellar Feedback:} This has both a thermal and a kinetic mode; thermal
feedback evaporates the cold phase within SPH particles and increases the temperature of
the hot phase, while kinetic feedback is modeled as a stochastic wind (as in
SH03) -- gas mass is lost due to
galactic winds at a rate $\dot{M}_w$, which is proportional to the star formation rate
$\dot{M}_{\ast}$, such that $\dot{M}_w = \eta\dot{M}_{\ast}$, with $\eta=2$. SPH particles
near the star forming region will be subjected to enter in the wind in an stochastic way.
Those particles impacted upon by the wind will be given an isotropic velocity kick of
$v_{\rm w}$ = 400 km/s and will freely travel without feeling pressure forces up to 20 kpc
distance from their original positions
\smallskip

\noindent \emph{SMBH Growth \& AGN Feedback:} These processes are not included.

\paragraph*{\bf Colour \& line style scheme} 

In all the radial plots below we distinguish codes including AGN feedback from codes which only
include stellar feedback. The first group is identified by dashed lines and the second one by solid lines.
Each code is identified by a different color. In all the plots, the codes are ordered by decreasing gas fraction
at $R_{500}^{crit}$ from left to right (or top to bottom).


\subsection{The Data}
We use zoom simulations of clusters produced with a variety of codes running full physics (FP) models, building upon the dark matter only and non-radiative simulations of S15. The initial conditions for our zoom simulations were drawn from the MUSIC-2 cluster catalog (\citealt{Sembolini2013}; \citealt{Sembolini2014}; \citealt{Biffi2014}) \footnote{Specifically, it is cluster 19 of MUSIC-2 dataset; all the initial conditions of MUSIC clusters are available at \texttt{http://music.ft.uam.es}} of re-simulated halos from the MultiDark cosmological simulation \footnote{A dark-matter only simulation containing 2048$^3$ particles in a (1$h^{-1}$Gpc)$^3$ cube performed using ART \citep{Kravtsov97} at the NASA Ames Research centre \cite[][]{Prada12}}. All the data from the parent simulation are accessible online through the {\itshape MultiDark Database}\footnote{www.cosmosim.org}. Our chosen cosmology corresponds to the best-fitting $\Lambda$CDM model to WMPA7+BAO+SNI data ($\Omega_{\rm m} =0.27$, $\Omega_{\rm b} = 0.0469$, $\Omega_{\Lambda}= 0.73$, $\sigma_8 =0.82$, $n = 0.95$, $h = 0.7$, \citealt{Komatsu11}). The effective resolution of these simulations is $m_{\rm DM}=9.01\times10^8\;h^{-1}$M$_\odot$ and, for the SPH codes, $m_{\rm gas}=1.9\times10^8\;h^{-1}$M$_\odot$.

The mass of a gas element naturally varies in our mesh codes. Star particle masses varies from code to code depending on how many generations of stars a gas element produces and the mass of the gas element being converted into a star particle.

All the halos were identified and analyzed using the Amiga Halo Finder, 
\ahf\ (\citealt{Gill04}; \citealt{Knebe09}; freely available from
\texttt{http://popia.ft.uam.es/AHF}).

\section{Bulk Properties} \label{sec:bulk}
Before we focus on the various components of our simulated clusters, we analyze the impact that the different subgrid models adopted in full physics simulations (FP) have on the bulk properties of the cluster.

\begin{figure}
   \includegraphics[width=0.495\textwidth]{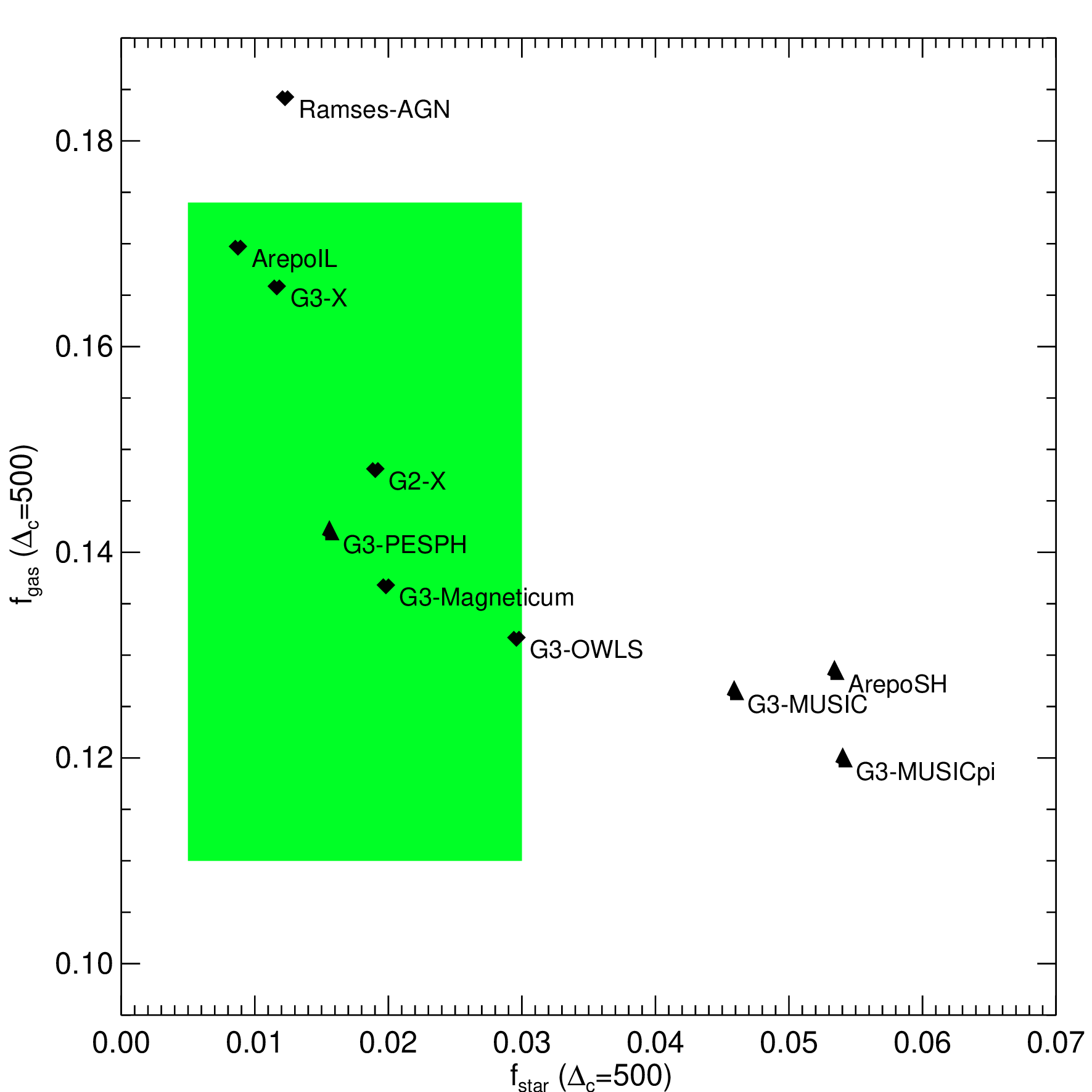} 
    \caption{Values of $f_{\rm gas}$ and $f_{\rm star}$ as calculated at $\Delta_c$ = 500 for the different codes. The green area corresponds to the phase space supported by observations. Codes including AGN feedback are represented as diamonds, codes not including AGN feedback as triangles. }
    \label{fig:fgas_fstar}
\end{figure}

As already mentioned in Section \ref{sec:introduction}, one of the main goals of modern simulations is to give a description of the baryonic (galaxies and ICM) component of clusters which succeeds in reproducing observational results.
We therefore start our analysis by testing how the different codes used in this work compare with measurements of the gas and stellar components as provided by observations.
We show in Figure~\ref{fig:fgas_fstar} the values of $f_{\rm gas}$ as calculated at $R_{500}^{\rm crit}$, the radius enclosing $\Delta_c$ = 500 times the critical density (the gas fraction with respect to the total mass of the cluster) against those of $f_{\rm star}$ (the star fraction) evaluated at the same overdensity. The green area indicates the range of values {\em allowed} by observations; as observational results still do not agree (e.g. \citealt{Gonzalez13} invokes higher gas fractions for massive clusters with respect to previous results, see Section \ref{sec:baryons} for a more detailed discussion), we set very non-restrictive limits to the extreme permitted values: 0.11 $< f_{\rm gas} <$ 0.174 (the value of the cosmic ratio according to WMAP7) and 0.005 $< f_{\rm star} <$ 0.03. We see that most of the codes not including AGN feedback show values of the stellar fraction which have been ruled out by observations, although they are able to reproduce the gas content. In this work we do not use an observational approach to estimate baryonic masses (e.g. measuring the gas fractions from synthetic X-rays observations), but we estimate the masses by simply counting the number of particles inside a fixed radius.

\begin{figure}
    \includegraphics[width=0.495\textwidth]{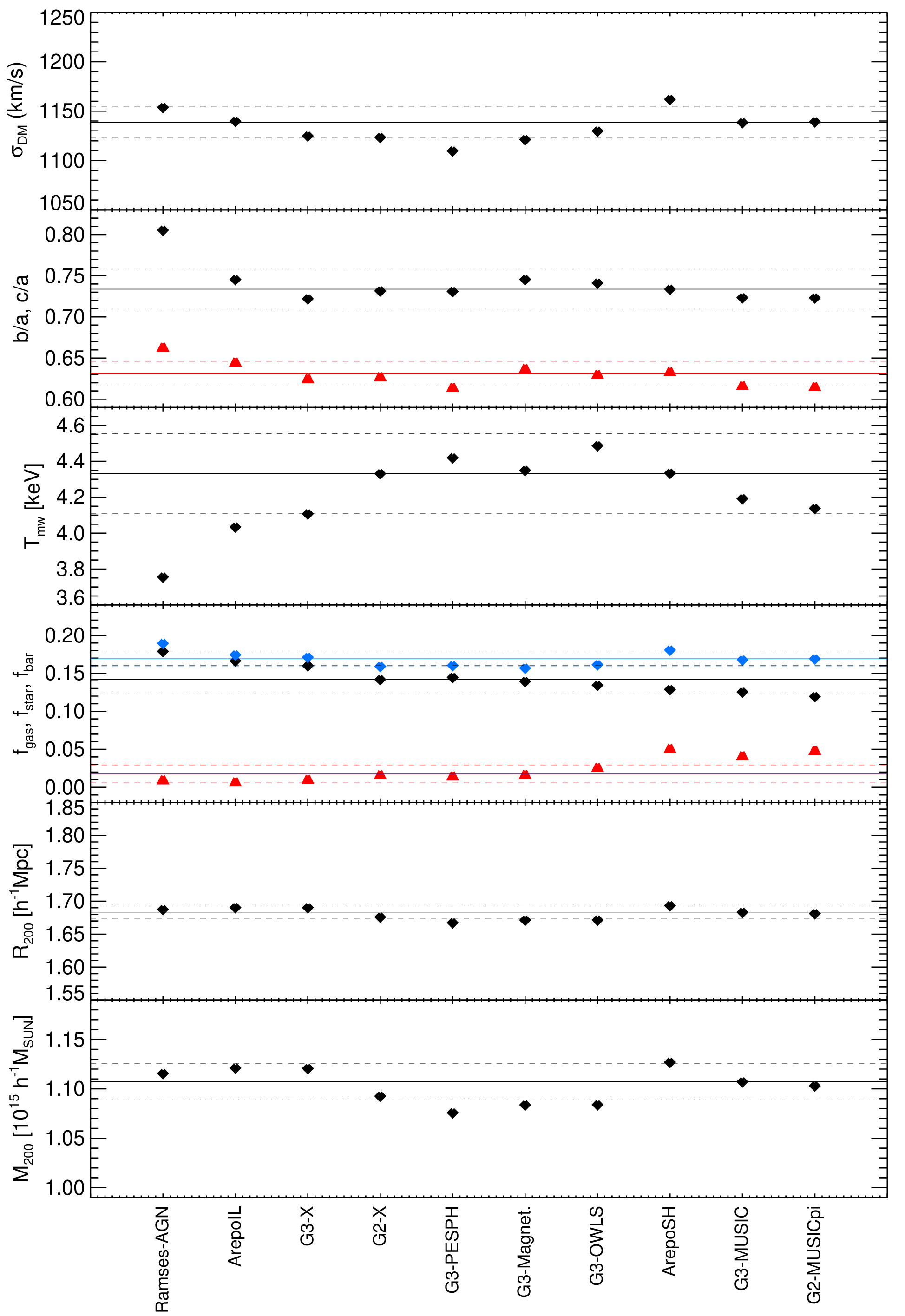}
    \caption{Global properties of the cluster produced by different codes. All quantities are computed within $R_{200}^{\rm crit}$. From top panel to bottom panel: $(1)$ the one-dimensional velocity dispersion of the dark matter, $(2)$  the axial ratio ($b/a$ in black, $c/a$ in red), $(3)$ the mass-weighted temperature,  $(4)$ the gas fraction (black), the star fraction (red) and the total baryon fraction (blue), $(5)$ the radius and $(6)$ the total cluster mass. The solid lines represent the median value for each one of the plotted quantities and the dashed lines $\pm$ the 1-$\sigma$ scatter.}
    \label{fig:global}
\end{figure}

Figure \ref{fig:global} shows a selection of global properties calculated within $R_{200}^{\rm crit}$, the radius enclosing 200 times the critical density: radius, mass, mass-weighted gas temperature, gas and stellar fractions, shape parameters (here we report the values of the minor semi-axes, $b$ and $c$, normalized to that of the major semi-axis, $a$) and the one dimensional velocity dispersion, {\it $\sigma_{\rm DM}$}. The first feature is that the scatter in FP simulations is higher than in the non-radiative (NR) case (see S15). The mean values for the total mass, radius, shape (with the exception in this case of \RamsesA) and DM velocity dispersion are extremely close to those in the non-radiative runs and still have very low scatter (less than 2 per cent). 

\par 
More importantly, pronounced differences lie in the baryonic sector. The temperature (4.3 keV, corresponding approximately to 5$\times$10$^7$K) is $\sim$20 per cent higher in FP simulations than in NR  models (3.7 keV) and has a scatter around 5 per cent compared to that of 2 per cent registered in the NR comparison. The gas fraction is lower than what was found in the non-radiative case (as some of the gas has been converted to stars), especially for the codes which do not include AGN feedback. The overall fractions show significant scatter: $f_{\rm gas}\sim0.12-0.18$ and a code-to-code scatter of 30-40 per cent; the discrepancies are more dramatic for the stellar component, where $f_{\rm star}$ varies between $0.01-0.05$ . The total baryon fraction ($f_{\rm bar} = f_{\rm gas} + f_{\rm star}$)  shows a more moderate scatter (around 10 per cent) and most of the codes show values around 0.16, very close to the cosmic ratio (here we adopt the value -used for our simulations- of $\Omega_b/\Omega_m \sim$ 0.174 reported using WMAP7+BAO+SNI data by \citealt{Komatsu11}). \RamsesA\ is the outlier, showing a baryon fraction that is slightly larger than the cosmic ratio ($f_{\rm bar} \sim$ 0.18).
Interestingly, we observe a trend in the AGN codes, from \RamsesA\ to \owls\: the temperature tends to increase and at the
same time, the gas fraction tends to decrease. This may suggest a variation
in feedback strength from left-to-right (as more and more gas is expelled,
the  remaining gas is hotter).

\begin{figure}
    \includegraphics[width=0.495\textwidth]{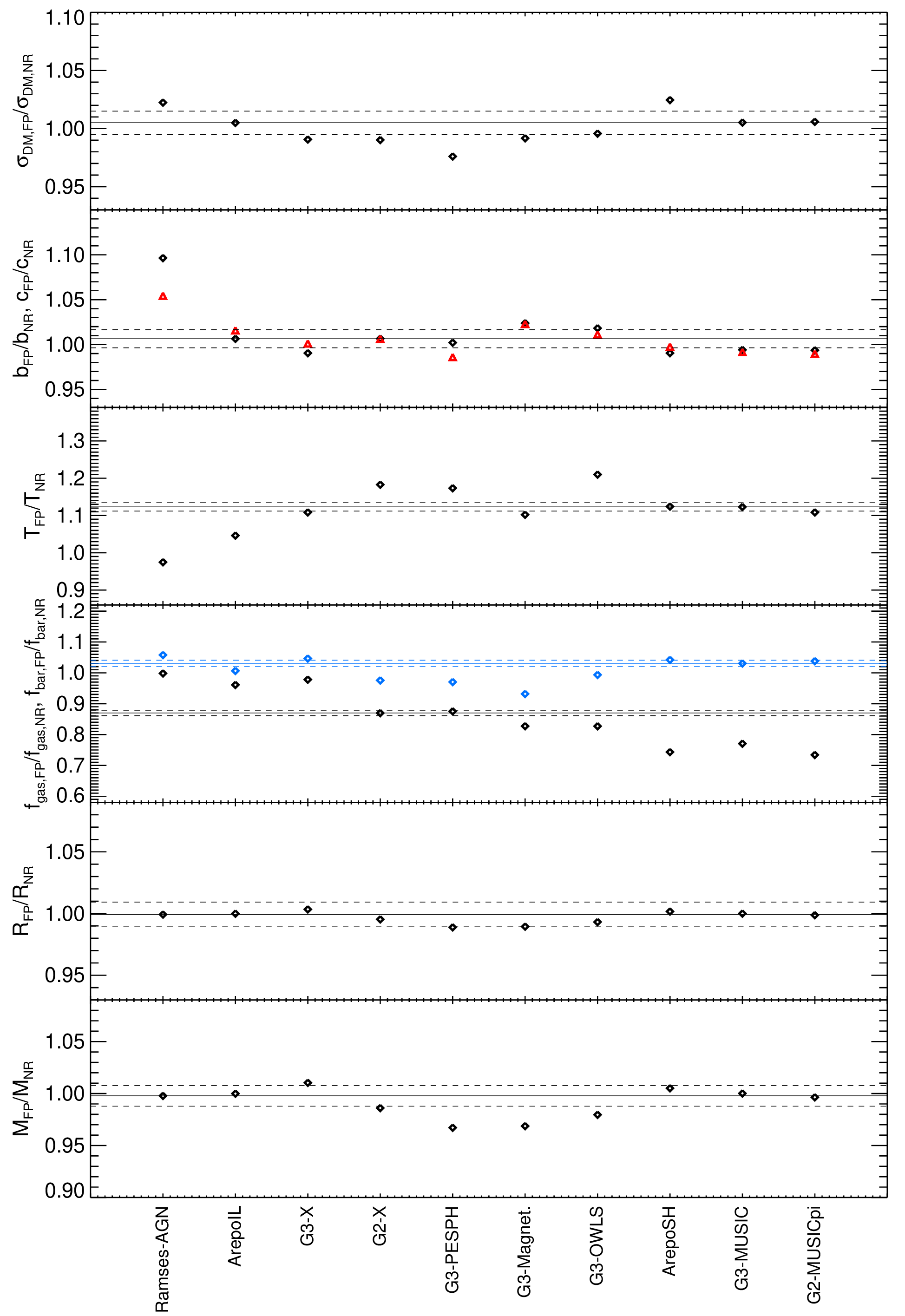}
    \caption{Ration between the same global properties shown in Figure \ref{fig:global} and the same values calculated for the correspondent NR runs and shown in Figure 4 of S15. The solid lines represent the median value for each one of the plotted quantities and the dashed lines $\pm1\%$.}
    \label{fig:ratio}
\end{figure}

Figure \ref{fig:ratio} shows how the main global cluster properties reported in Figure \ref{fig:global} changed in full physics simulations with respect to the NR runs reported in S15. The quantities that exhibit less scatter (e.g. mass and radius) are, as expected, also the ones whose values were basically unchanged with respect to the NR models, with differences lower than 1 per cent (only for \RamsesA\ some of these values are 5 per cent higher than its NR version) and scatters between 1 and 3 per cent. The temperature and gas fraction, which depend only on the baryon component and are therefore more affected by radiative processes, exhibit higher differences:  as the gas is heated by the different energy injection mechanisms included in the FP simulations, temperatures are on average 10 per cent higher (with the only exception of \RamsesA, which registers a temperature a few per cent lower than its NR model) with a scatter of 7 per cent. Furthermore, as part of the baryon component is now converted into stars, the gas fraction is now substantially lower: we find a median value of 15 per cent and a scatter of 13 per cent. On the other hand, the methods with the lowest portion of baryons converted into stars (see Section \ref{sec:stars}), such as \RamsesA\ and \gx, show a gas fraction very close to the value registered for the corresponding NR version. The total baryon fraction is either almost unaltered or 5-10 per cent lower than in the NR case for almost all the codes.

\section{Dark matter} \label{sec:DM}
\begin{figure*}
    \centering
    \begin{tabular}{ccc}
     \includegraphics[width=50mm]{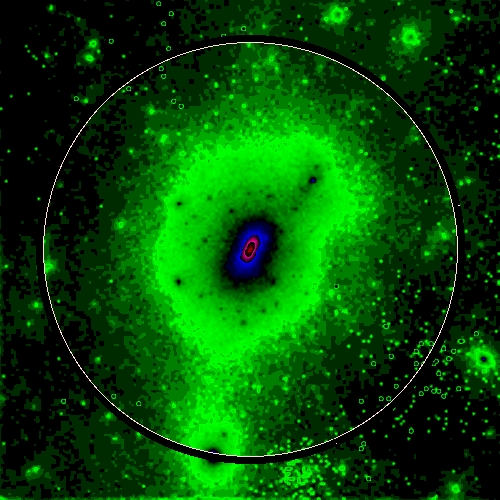}&
      \includegraphics[width=50mm]{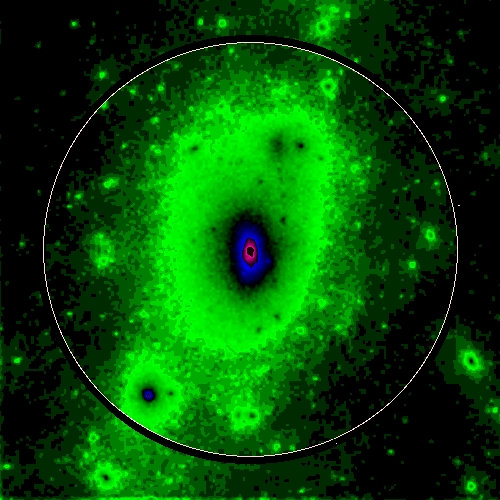}&
      \includegraphics[width=50mm]{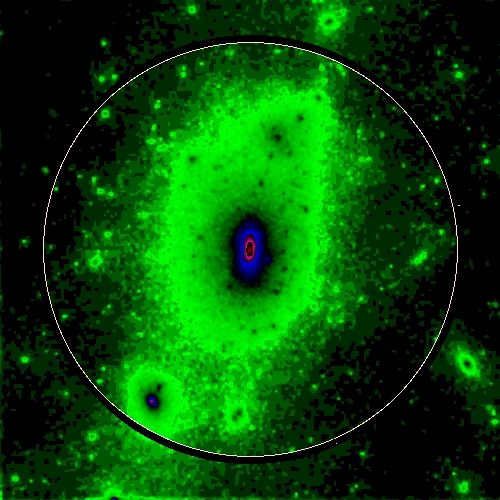}\\
  \RamsesA\ & \arepoIL\ & \gx\\
  
       \includegraphics[width=50mm]{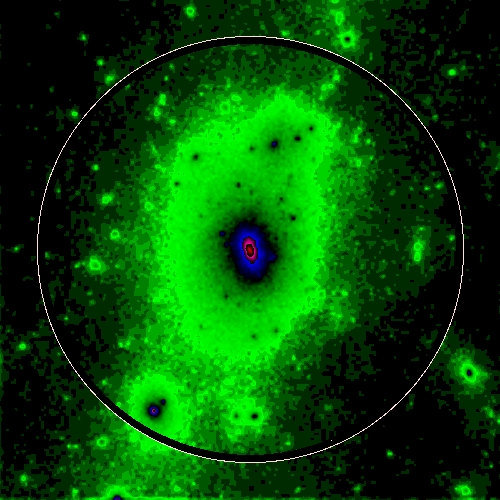}&       
        \includegraphics[width=50mm]{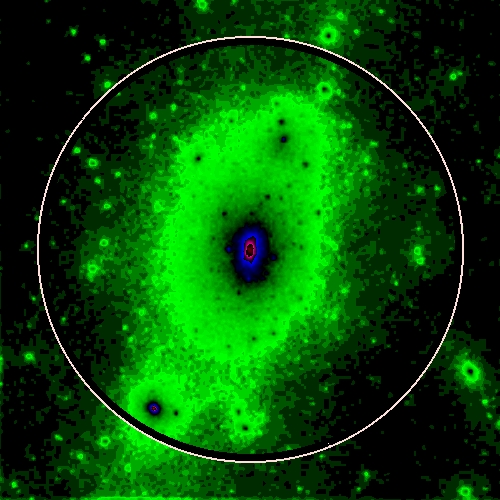}&
        \includegraphics[width=50mm]{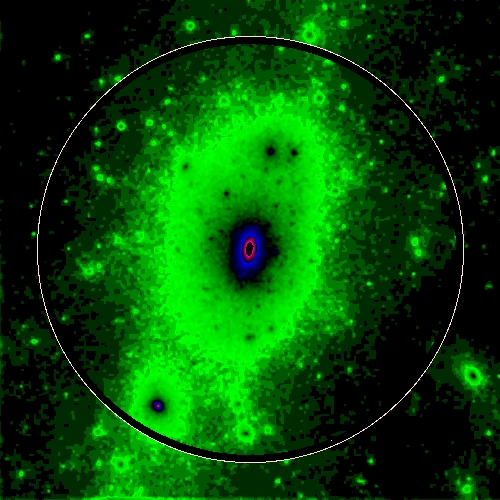}\\
  \gxx\ & \pesph\ & \magneticum\\
       \end{tabular}
     \begin{tabular}{cc}
       \includegraphics[width=50mm]{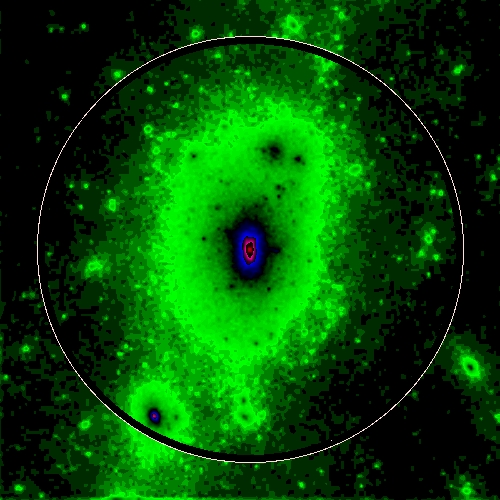}&
        \includegraphics[width=50mm]{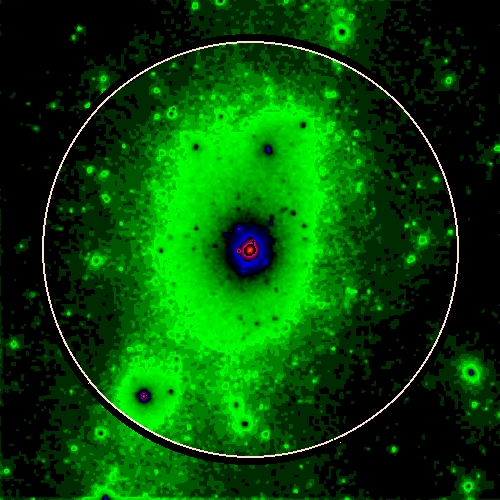}\\     
 	\owls\ & \arepoSH\\
        
        \includegraphics[width=50mm]{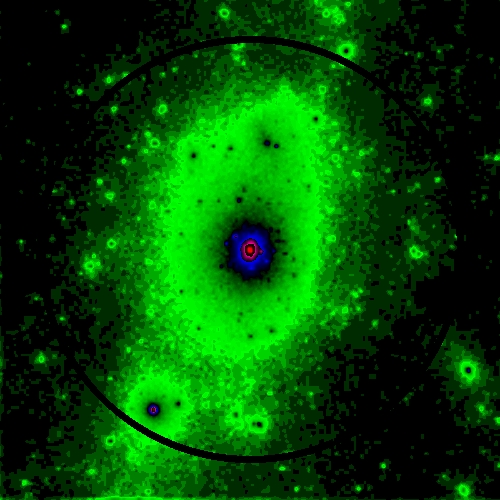}&
        \includegraphics[width=50mm]{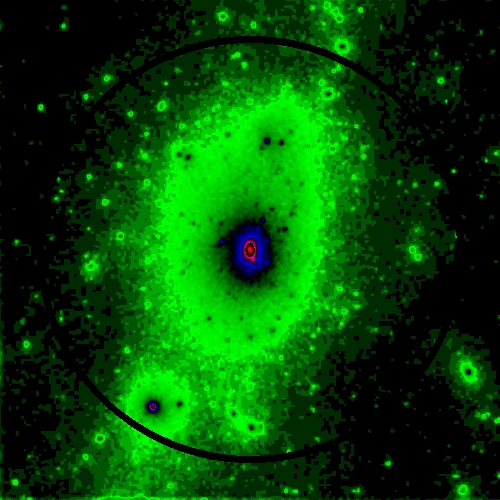}\\
        \music\ & \musicP\\
    \end{tabular}
    \includegraphics[width=70mm]{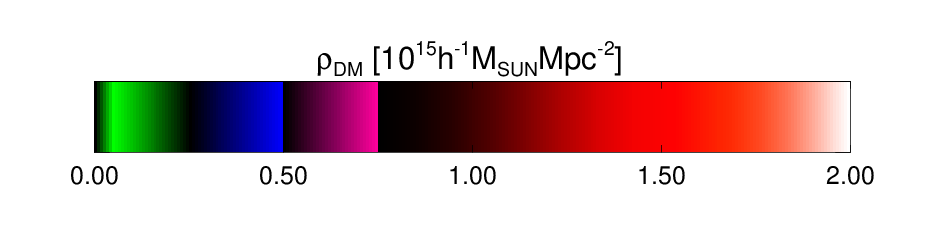} 
    \caption{Projected dark matter density at $z=0$ for each simulation as indicated. Each box is 2$h^{-1}$Mpc  on a side. The white circle indicates $M_{200}^{\rm crit}$ for the halo, the black circle shows the same but for the \music\ simulation.}
    \label{fig:DM_maps}
\end{figure*}

\begin{figure}
    \includegraphics[width=0.495\textwidth]{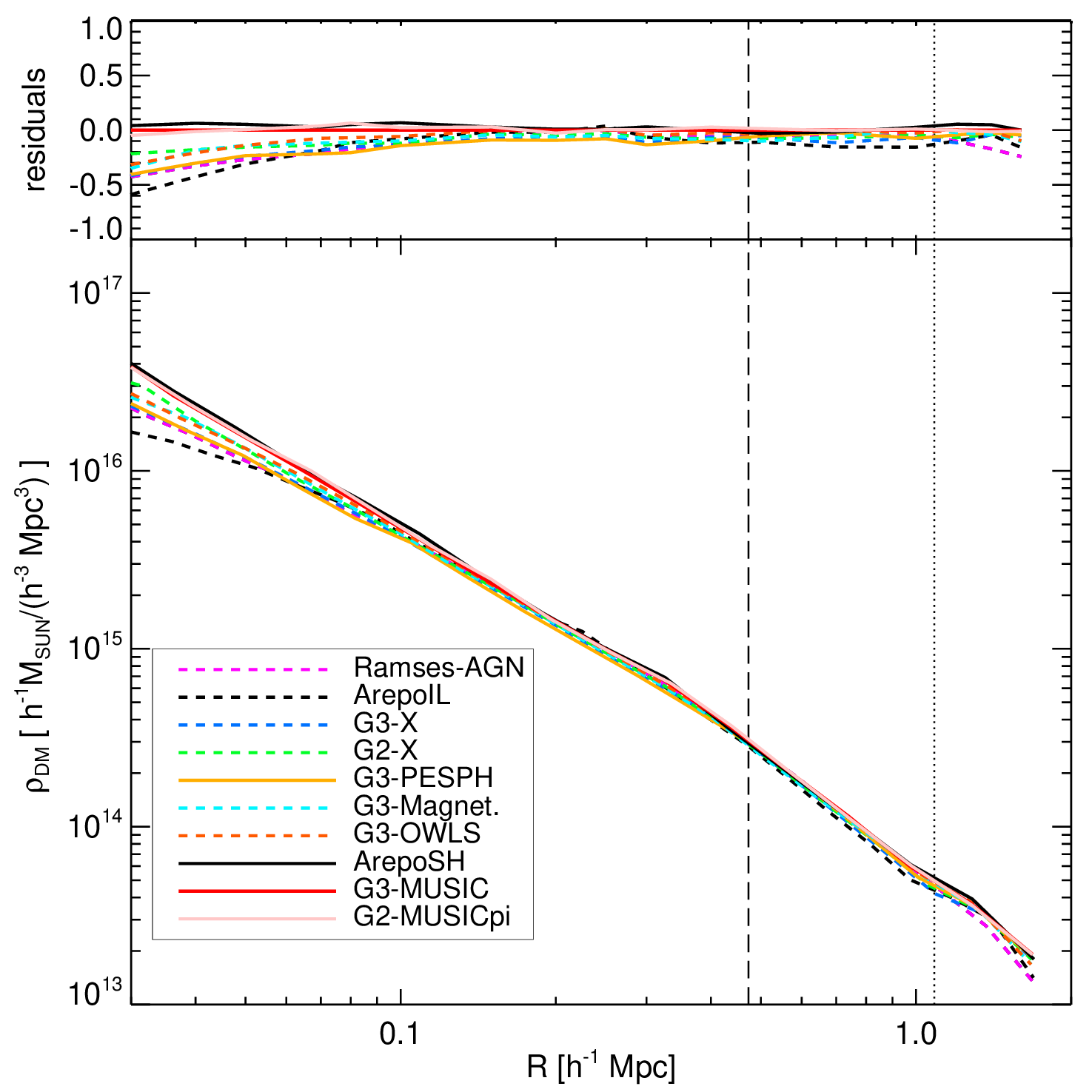}
    \caption{Radial density profiles at $z=0$ (bottom panel) and difference between each listed simulation and the reference \music\ (top panel). The dashed line corresponds to $R^{\rm crit}_{2500}$ and the dotted line to $R^{\rm crit}_{500}$ for the reference \music\ values.}
    \label{fig:rho_dm}
\end{figure}

A visual comparison of the density field centered on the cluster at $z = 0$ is presented in Figure \ref{fig:DM_maps} and density profiles are shown in Figure \ref{fig:rho_dm}. Although all the codes successfully recover the same object and its main features (e.g. the position of the main subhalo, which in the maps is located  at 7 o'clock close to $R^{\rm crit}_{200}$, except for \RamsesA, which seems to have a slightly different merger phase), the dark matter distribution differs significantly more than what was found in S15 for the dark matter-only and non-radiative models. 

These differences in the dark matter distribution arise in response to the baryons. As baryons cool they can pull in dark matter with an effect similar to adiabatic contraction (\citealt{Eggen62}; \citealt{Zeldovich80}). This contraction may look surprising at first sight as dark matter dominates the mass budget of the cluster, exceeding baryonic matter by a factor of $\sim$ 6. However, the gravitational field in the central regions of a halo is dominated by stars, which formed from the condensations of cooling baryons. The amount of the contraction was studied for the first time in cosmological simulations by \cite{Gnedin04} (and recently revisited by \citealt{Capela14}). These studies indicated that cooling and star formation can produce clusters and galaxies with central dark matter densities that are an order of magnitude higher than analogues in non-radiative runs. \cite{Duffy10} studied the effects of feedback from star formation and AGN, finding large variations and much less contraction when AGN feedback is included.

\par 
Of greater significance is the variety in the dark matter distributions, most easily seen in the radial profiles of Figure \ref{fig:rho_dm}. The first notable systematic is that codes which exhibit a stronger contraction are those which {\em do not} include AGN feedback (\music, \musicP, \arepoSH), with the exception of \pesph. These codes have inner regions ($R < 100h^{-1}$kpc) with densities a factor of 2 higher than the other codes. Many studies show that simulations of clusters that lack a physical mechanism to stop the central cooling of the gas are affected by the problem of overcooling (e.g. \citealt{Suginohara98}; \citealt{Lewis00}; \citealt{Tornatore03}; \citealt{Nagai04}). These codes have a notably higher fraction of the baryons in the form of cold gas and stars within the virial radius than inferred from observations, 30-50 per cent vs 10-20 per cent, and are expected to produce more stars (see Section \ref{sec:stars} for a more detailed discussion). This picture fits with their higher dark matter concentrations. 

\par 
Codes that include AGN feedback do not have such a pronounced contraction, with dark matter profiles similar to that reported for NR runs (see Figure 2 in S15). The interesting exception noted before is \pesph, which has a profile similar to \gxx\ and \gx. Among the AGN codes, \arepoIL\ experiences the smallest contraction, a factor of 2 less than the other codes. As the contraction is related to the star formation efficiency, it is no surprise to find that \arepoIL\ is one of the codes with the fewest stars (see Figure \ref{fig:fstar} in Section \ref{sec:stars}). 

\par 
The profiles not only show systematic differences, the code-to-code scatter in full physics simulations is considerably higher (up to a factor of 5 between the two different versions of \arepo\ at the center of the halo) than that observed in the DM-only and non-radiative runs (see Figures 1, 2 and A1 of S15), where differences never exceeded 20 per cent. This scatter occurs primarily in the central regions. The cluster outskirts show a scatter of $\lesssim$ 10 per cent. The large difference between the two different versions of \arepo\ confirms how the dark matter distribution depends on the subgrid physics adopted, and in particular by how energy is injected into the gas reservoir.

\section{Baryons} \label{sec:baryons}
\begin{figure*}
    \centering
    \begin{tabular}{ccc}
     \includegraphics[width=50mm]{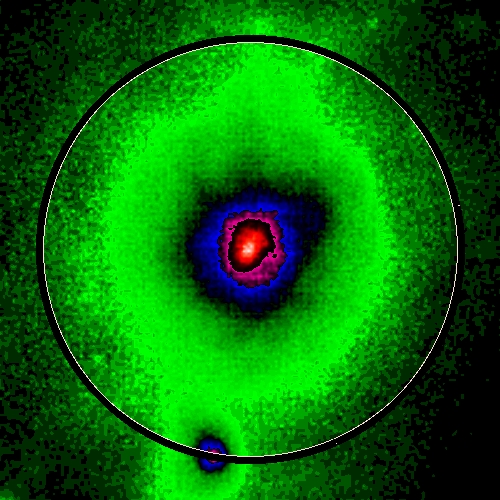}&
      \includegraphics[width=50mm]{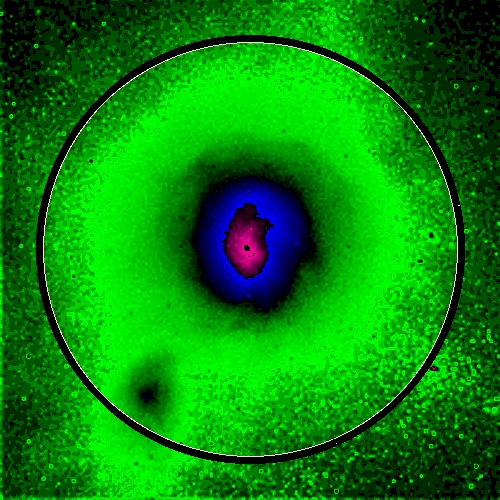}&
      \includegraphics[width=50mm]{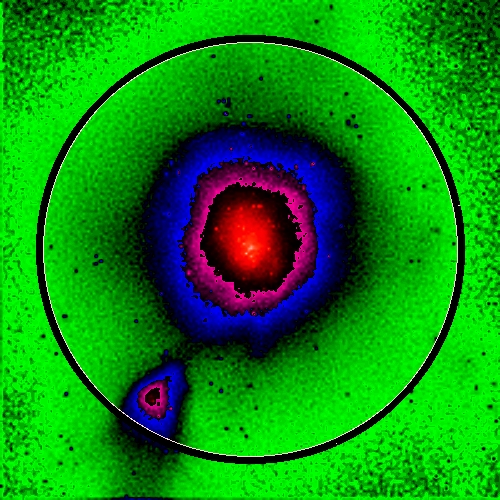}\\
  \RamsesA\ & \arepoIL\ & \gx\\
 
       \includegraphics[width=50mm]{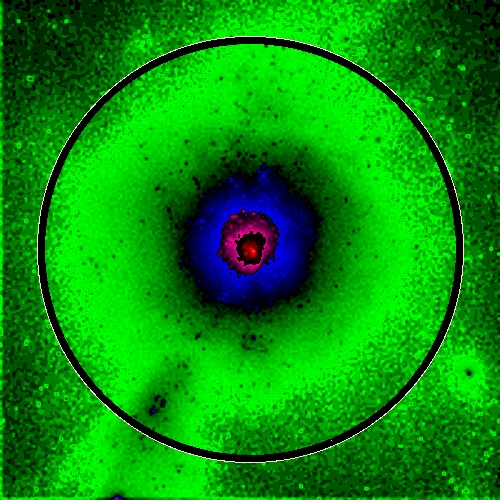}&       
        \includegraphics[width=50mm]{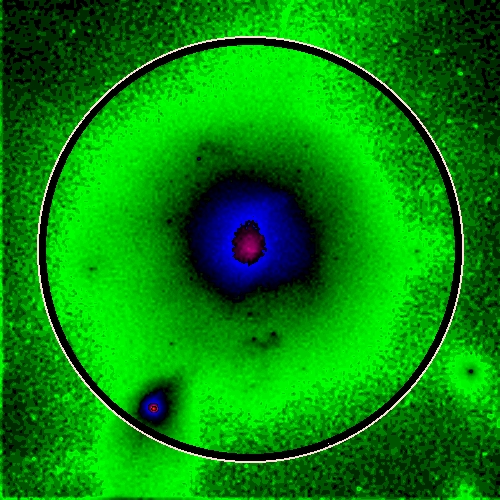}&
        \includegraphics[width=50mm]{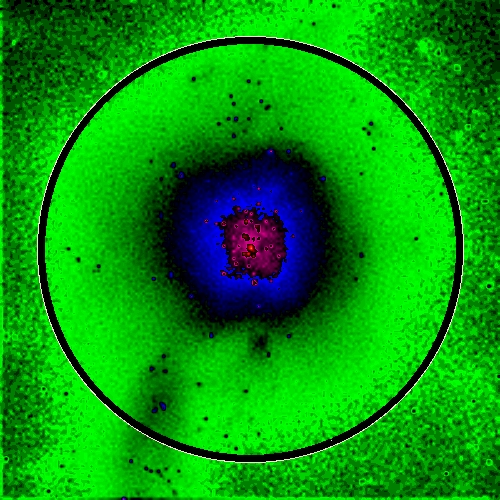}\\
  \gxx\ & \pesph\ & \magneticum\\
     \end{tabular}
     
     \begin{tabular}{cc}    
       \includegraphics[width=50mm]{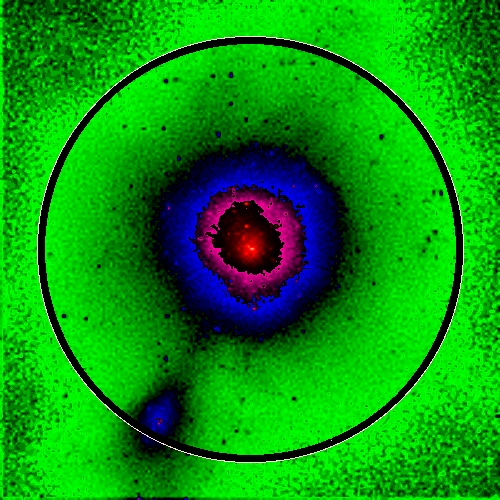}&
        \includegraphics[width=50mm]{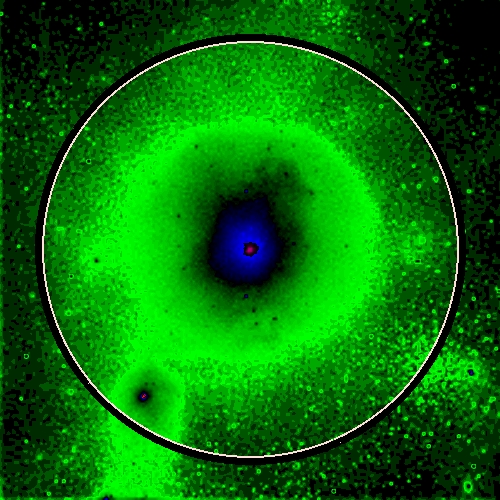}\\     
  \owls\ & \arepoSH\\
        
        \includegraphics[width=50mm]{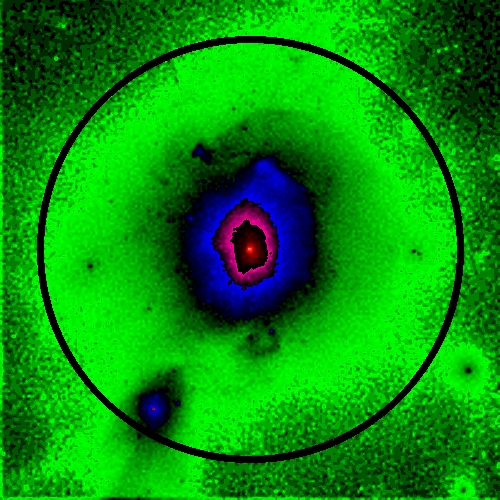}&
        \includegraphics[width=50mm]{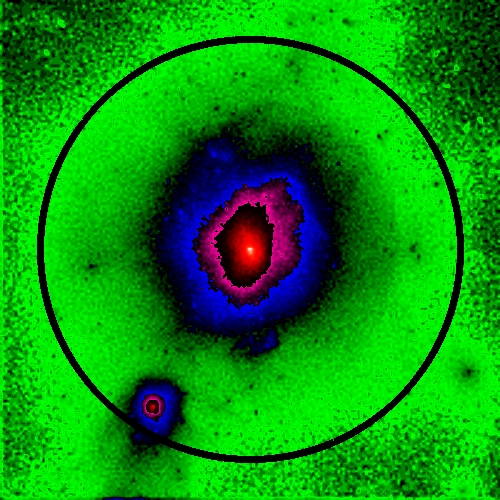}\\
       \music\ & \musicP\\
    \end{tabular}
    \includegraphics[width=70mm]{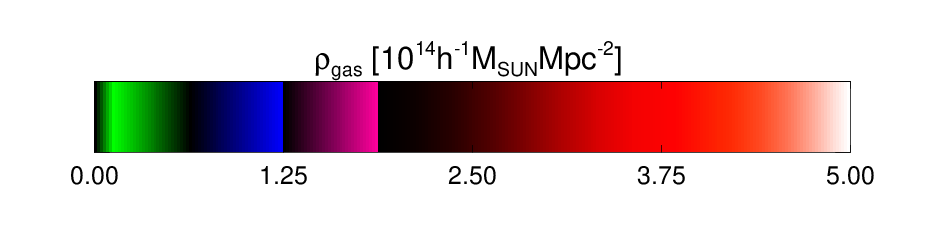} 
    \caption{Projected gas density at $z=0$ for each simulation as indicated. Each box is 2$h^{-1}$Mpc  on a side. The white circle indicates $M_{200}^{\rm crit}$ for the halo, the black circle shows the same but for the \music\ simulation.}
    \label{fig:gas_maps}
\end{figure*}

\begin{figure*}
    \centering
    \begin{tabular}{ccc}
     \includegraphics[width=50mm]{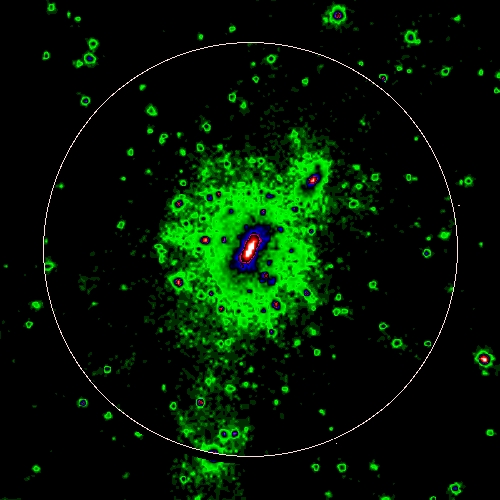}&
      \includegraphics[width=50mm]{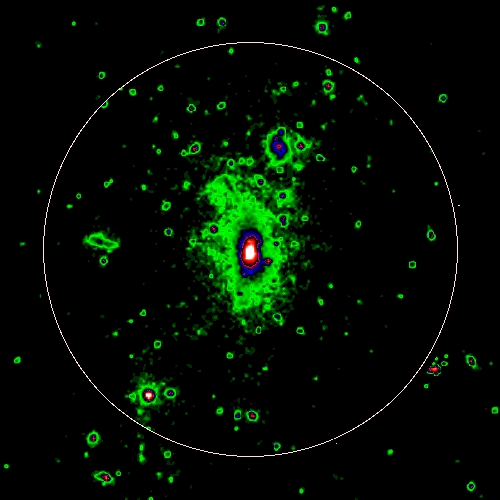}&
      \includegraphics[width=50mm]{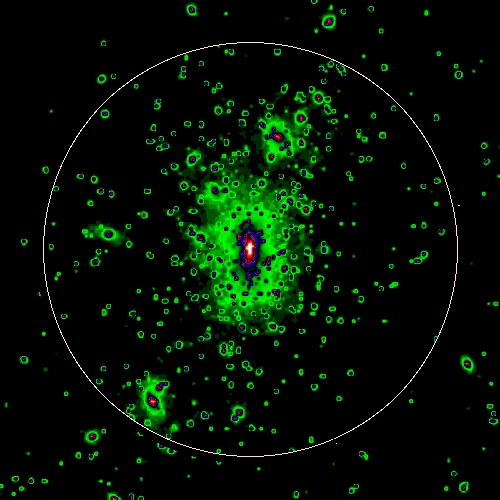}\\
  \RamsesA\ & \arepoIL\ & \gx\\
  
       \includegraphics[width=50mm]{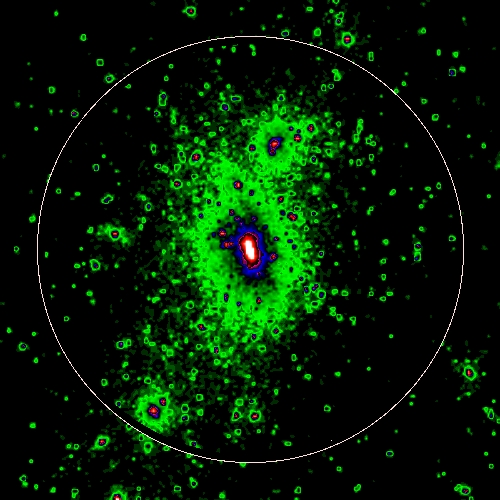}&       
        \includegraphics[width=50mm]{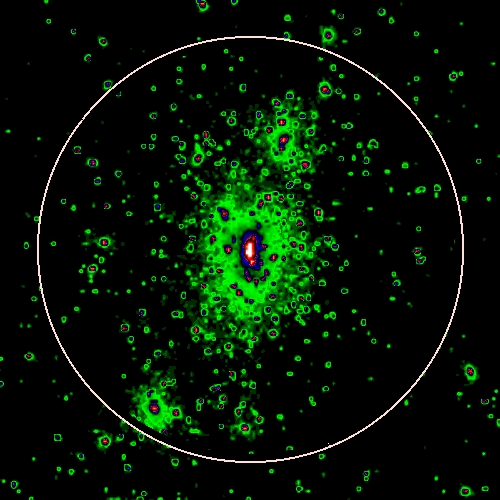}&
        \includegraphics[width=50mm]{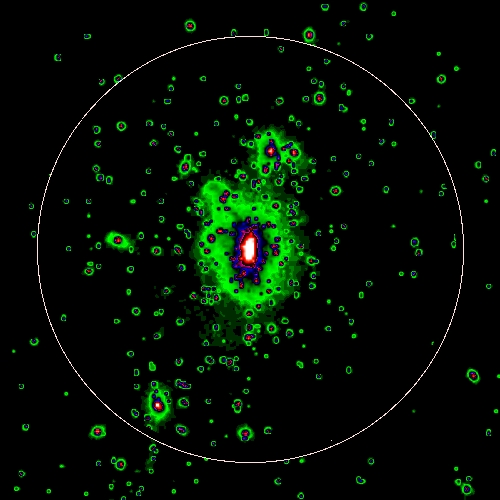}\\
   \gxx\ & \pesph & \magneticum\ \\
 \end{tabular}
 
  \begin{tabular}{cc}
       \includegraphics[width=50mm]{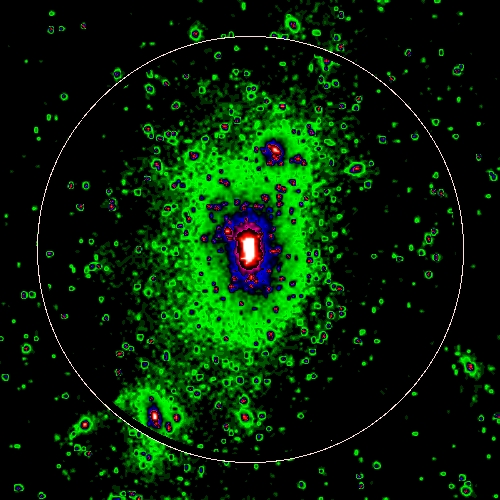}&
        \includegraphics[width=50mm]{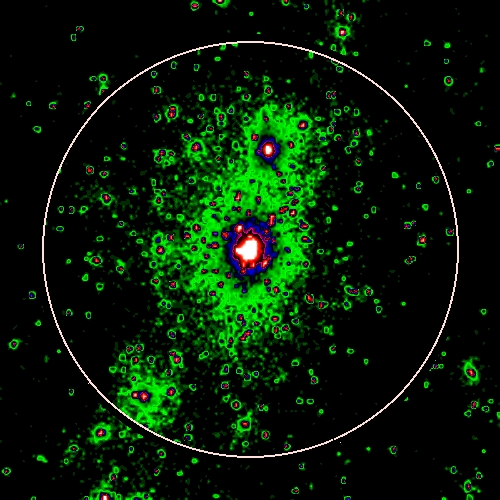}\\     
    \owls\ & \arepoSH\\
        
        \includegraphics[width=50mm]{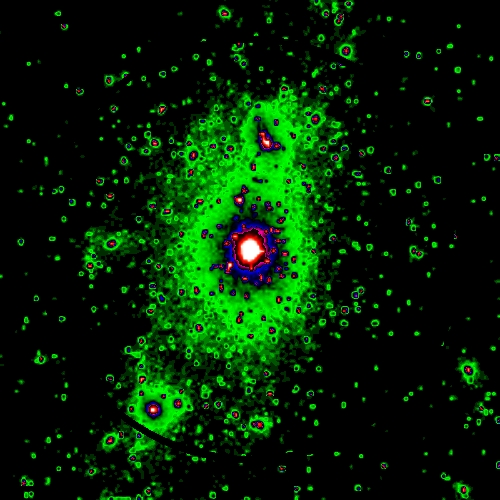}&
        \includegraphics[width=50mm]{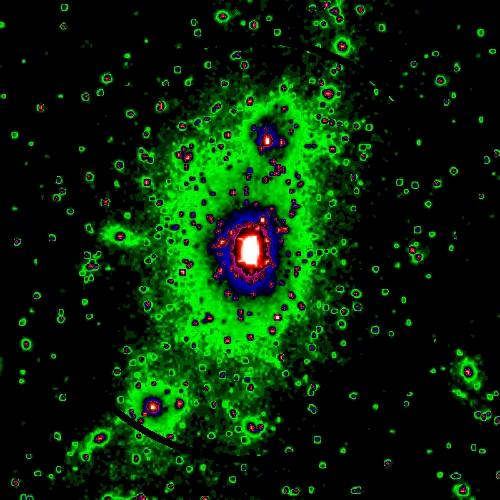}\\
        \music\ & \musicP\\
    \end{tabular}
    \includegraphics[width=70mm]{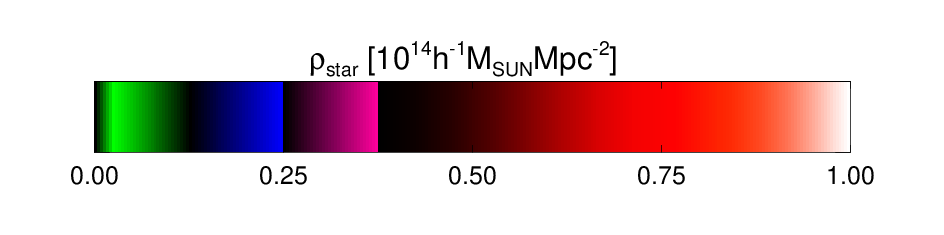} 
    \caption{Projected stellar density at $z=0$ for each simulation as indicated. Each box is 2$h^{-1}$Mpc  on a side. The white circle indicates $M_{200}^{\rm crit}$ for the halo, the black circle shows the same but for the \music\ simulation.}
    \label{fig:star_maps}
\end{figure*}

We now focus on the baryons in our simulated clusters. We show the $z = 0$ gas and stellar distributions of some relevant cluster properties produced by each code in Figures~\ref{fig:gas_maps}-\ref{fig:fstar}. 

\subsection{Gas}\label{sec:gas}
A visual comparison of the gas density field centered on the cluster at $z = 0$ is presented in Figure~\ref{fig:gas_maps}. There is a substantial amount of variation in the central gas density, with some methods (\arepoIL, \gx, \musicP, \owls) having significantly larger extended nuclear regions. Some codes appear to show numerous small dense gas clumps in the cluster outskirts, especially those including AGN feedback: in this case AGN prevents gas from cooling and forming stars, and therefore more gas is left in these substructures. More significantly, we observe that different subgrid physics applied to the same code (\arepo) produces very different gas environments. 

\par 
Figure~\ref{fig:star_maps} allows a visual comparison of stellar density distributions. The projected stellar densities appear to show even more variation. \RamsesA, \arepoIL\  and \gx\ have dense stellar objects whereas \arepoSH, \music\ and \musicP\ have significantly more extended stellar distribution. \owls\ also has an extended intra-cluster stellar halo but also has numerous stellar concentrations. Moreover, features of the gas distribution do not map to features in the stellar distribution, i.e., an extended gas distribution does not necessarily produce an extended stellar distribution. For instance, both \gxx\ and \RamsesA\ show a very high gas concentration in the core, but the latter produces a much more limited star distribution.

\begin{figure}
    \includegraphics[width=0.495\textwidth]{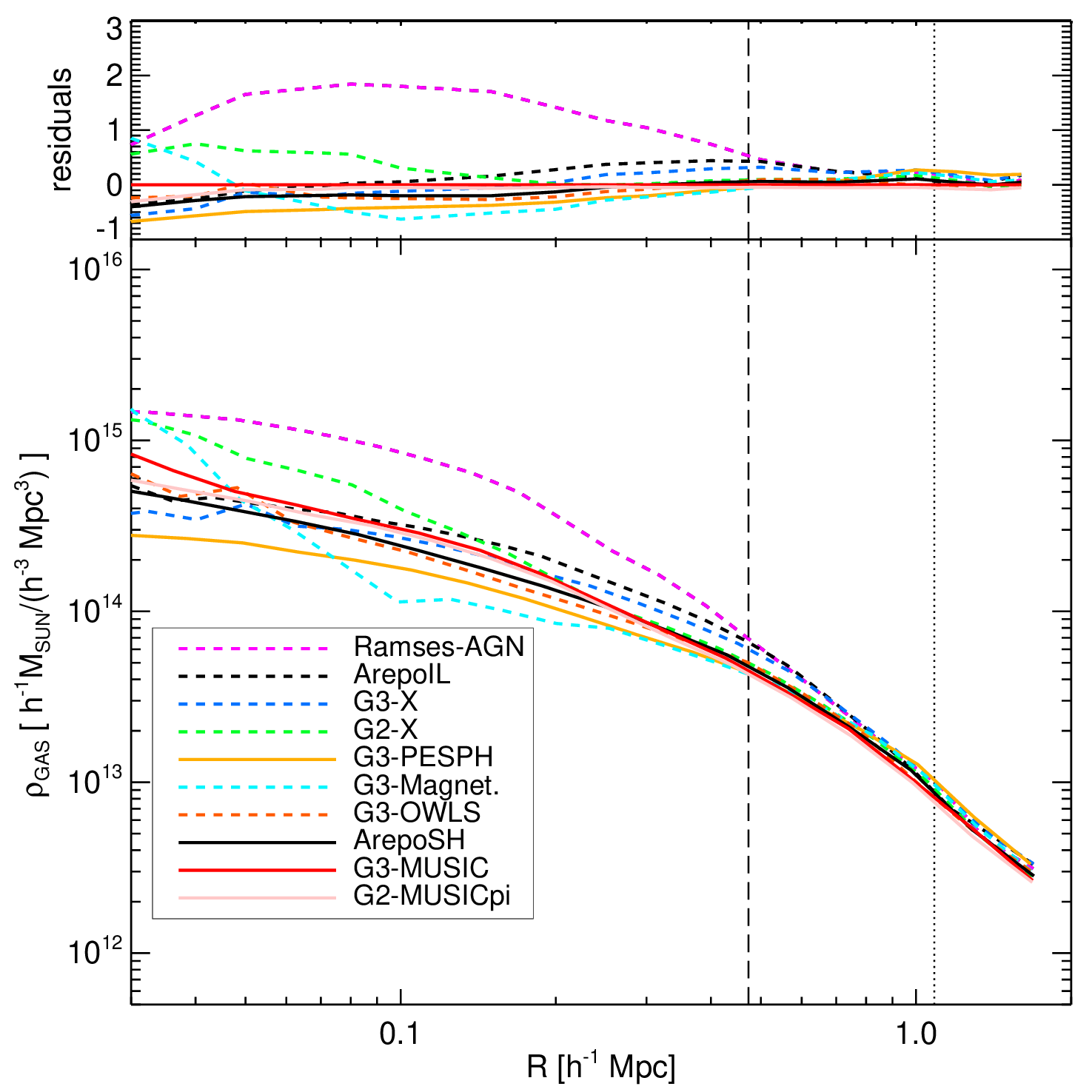}
    \caption{Radial gas density profile at $z=0$ (bottom panel) for each simulation as indicated and difference between each simulation and the reference G3-MUSIC simulation (top panel). The dashed line corresponds to $R^{\rm crit}_{2500}$ and the dotted line to $R^{\rm crit}_{500}$ for the reference G3-MUSIC values.}
    \label{fig:rho_gas}
\end{figure}

The gas differences seen in Figure~\ref{fig:gas_maps} are also evident in the radial gas density profiles presented in Figure~\ref{fig:rho_gas}. The code-to-code scatter in the central regions is $\sim$ 40 per cent and decreases in the outskirts of the cluster. The outliers are \pesph, which produces the lowest central density in the core (a factor of $\gtrsim3$ times smaller), and \RamsesA, which has the highest. In the outskirts the differences among codes are much more contained at overdensities lower than 2500 (although \RamsesA, \arepoIL\ and \gx\ show slightly higher gas densities). Interestingly, we also notice that \music\ and \arepoSH, which adopt the same star formation model (SH03), show very similar gas fraction profiles in the outskirts. 
The scatter is generally higher than in the non-radiative case (see Figure 6 of S15).
As anticipated visually by Figure~\ref{fig:gas_maps}, the same hydrodynamics code with different subgrid physics produces different gas distributions (e.g. \arepo). Furthermore, in the behavior of the gas density there is not a clear distinction between grid-based and modern SPH codes on the one hand and classic SPH on another hand as highlighted in the NR case (Figure 6 of S15).

\par
\begin{figure}
    \includegraphics[width=0.5\textwidth]{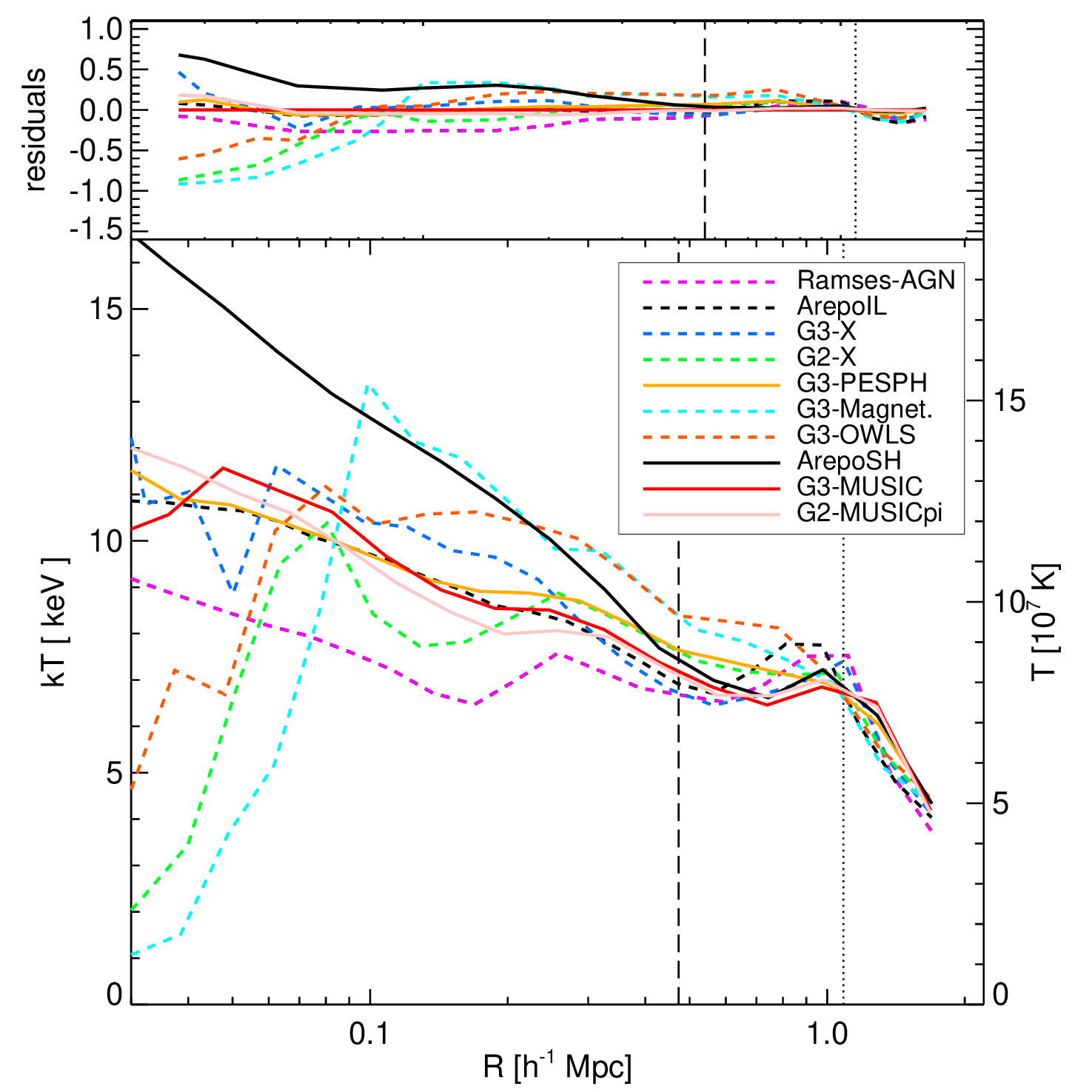}
    \caption{Radial temperature profile at $z=0$. Format similar to Fig.~\ref{fig:rho_gas}.}
    \label{fig:Tgas}
\end{figure}

\smallskip
We next show in Figure~\ref{fig:Tgas} the radial mass-weighted temperature profiles, defined as:
\begin{equation}
    T_{\rm mw} = \frac{\sum_i T_im_i}{\sum_im_i},
\end{equation}\label{eq:Tmw}
where $m_i$ and $T_i$ are the mass and temperature of the gas particles/cells. The code-to-code scatter is large, especially at the center of the cluster. \owls, \magneticum, \RamsesA\ and \gxx\ show a central temperature inversion, similar to that observed in non-radiative, classic SPH simulations: the inner temperature is $2-3$ times smaller than the peak value, which here is $8-10$~keV. In particular, \magneticum\ shows a very sharp temperature inversion at R $\sim$ 0.1$h^{-1}$Mpc: this effect is probably due to overcooling, as a large portion of the gas in the core is converted into a massive gaseous BCG.
In contrast, all the other codes display rising profiles going towards the the core, a behavior that is observed in modern SPH and mesh-based non-radiative simulations (see Figure 7 of S15). The typical peak temperature for this cluster in these codes is $10-13$ keV. The outlier amongst the codes with no temperature inversion is \arepoSH, which has an inner temperature exceeding 20~keV. Intriguingly, pronounced differences between codes including and not including AGN feedback are {\it not visible}. It is also interesting that some classic SPH codes (such as \music), which in the non-radiative simulations produce a central temperature inversion, now produce monotonically rising temperature profiles  (in agreement with \citealt{Rasia14}, which pointed out that radiative processes decrease the tension in temperature profiles between classic SPH and adaptive-mesh codes).

\par
\begin{figure*}
    \includegraphics[width=0.7\textwidth]{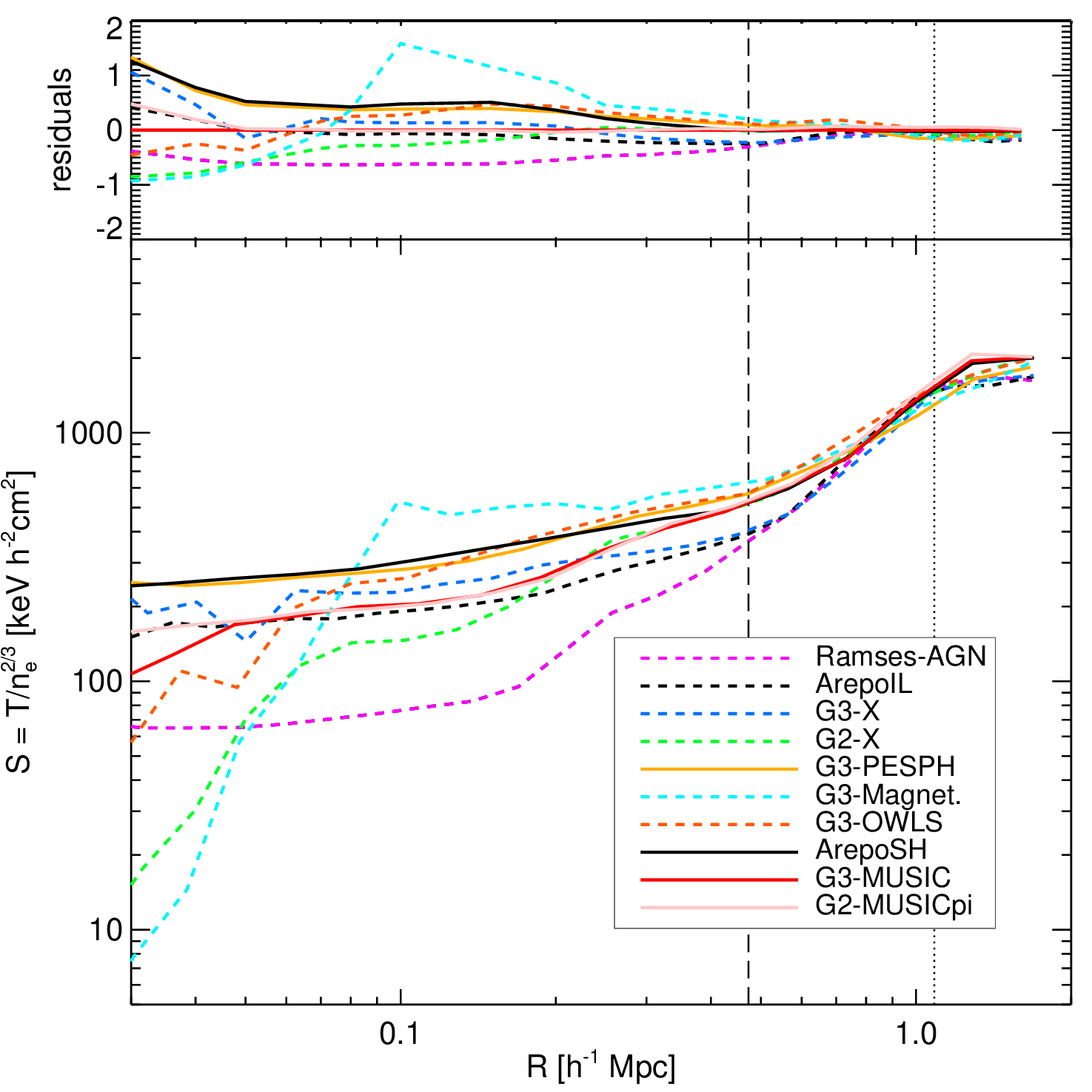}
    \caption{Radial entropy profile at $z=0$. Format similar to Fig.~\ref{fig:rho_gas}.}
    \label{fig:entropy}
\end{figure*}

\smallskip
We combine the gas density and temperature to produce the radial gas entropy profiles shown in Figure~\ref{fig:entropy}, where we adopt the definition of entropy commonly
used in the observational X-ray literature: 
\begin{equation}
    S(R) = \frac{kT_{\rm gas}(R)}{n_{\rm e}^{2/3}(R)}, 
\end{equation}
where $n_{\rm e}$ is the number density of free electrons of the gas. We observe that the differences between modern and classic SPH methods that had been displayed for the non-radiative case (see Figure 8 of S15) have been washed away to a certain extent with the inclusion of radiative subgrid physics. Radiative processes dominate the effect that different treatments of artificial viscosity and entropy dissipation have on the entropy profile. That is not to say that codes produce the same profile. Codes with temperature inversions (\magneticum, \gxx) still stand out. {\it However, the key result is that classic SPH codes such as \music\ and \owls\ no longer produce declining entropy profiles with decreasing radius: they now exhibit an almost-flat entropy core.} The other classic SPH code, \gxx, still displays a falling entropy inner profile. Subgrid physics {\em can} wash away the differences between classic SPH and mesh codes. Interestingly, the modern SPH code \pesph, which produced a falling inner entropy profile more similar to classic SPH in NR simulations than to other modern SPH methods, is now indistinguishable from the \arepoSH\ entropy profile. We also note that the introduction of radiative physics in the mesh code \arepo\ has pushed the entropy profile in the opposite direction. In non-radiative runs, \arepo\ produces flat entropy cores but it now has a shallow slope in both its subgrid versions. The grid-based code \ramses\ shows an almost flat entropy core, although significantly lower than some classic SPH codes such as \music. Another key result is that AGN feedback does not seem to play a dominant role in governing the entropy profile (e.g. the \music\ and \gx\ entropy profiles are similar). In general, codes produce an ``almost-flat'' central entropy profile, matching the observed overall X-ray profiles (see e.g. \citealt{Walker12}).
X-ray observations show flat entropy cores for NCC clusters and declining entropy profiles for CC clusters (see also \citealt {Rasia15} for a discussion on the effect of AGN $vs$ artificial diffusion on the entropy profiles and their relative importance in establishing the cool-coreness of clusters).
\par

\begin{figure}
    \includegraphics[width=0.495\textwidth]{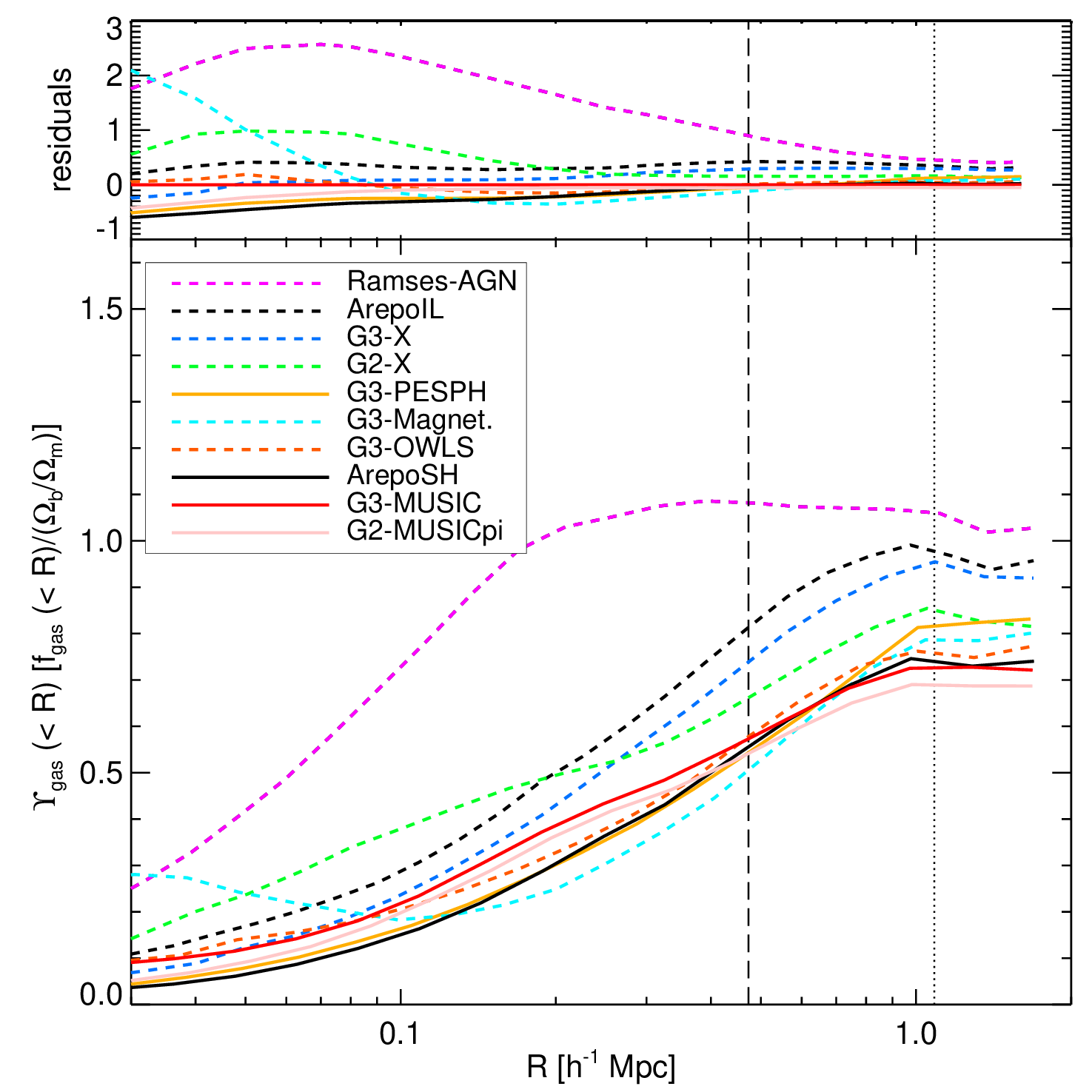}
    \caption{Cumulative radial gas fraction profile at $z=0$. Format similar to Fig.~\ref{fig:rho_gas}.}
    \label{fig:fgas}
\end{figure}

\smallskip
A natural follow-up question to ask is whether similar (dis)agreement between codes is seen for the gas fraction (see Figure \ref{fig:fgas}):
\begin{equation}
    \Upsilon_{\rm gas} = \left[\frac{M_{\rm gas}(<R)}{M(<R)}\right]\left(\frac{\Omega_b}{\Omega_m}\right)^{-1}.
\end{equation}
Another key result of S15 was that classic SPH codes typically have very baryon rich cores, $\Upsilon_{\rm gas}$(R$ < $0.1h$^{-1}Mpc$) $\gtrsim$0.4, whereas newer SPH schemes, mesh codes and \arepo\ produce cores with $\Upsilon_{\rm gas}(R<0.1h^{-1}{\rm Mpc})\lesssim0.2$. Full physics simulations contain little gas in the central regions as a result of star formation, regardless of the code used.  \RamsesA, \arepoIL, \gxx\ and \gx\ show a gas fraction that is significantly higher than for the other codes at $R^{\rm crit}_{2500}$ and, in the case of \RamsesA, it exceeds the cosmic ratio outside $R^{\rm crit}_{2500}$: as shown in Figure \ref{fig:ratio}, its value at $R^{\rm crit}_{200}$ is even higher than in the NR case. 

The key systematic difference between codes arises from AGN feedback, which produces in \RamsesA, \arepoIL\ and \gx\ the most evident effect (with the last two showing very similar results). AGN feedback {\em increases} gas fractions throughout the cluster with respect to radiative runs with no AGN, especially outside $R^{\rm crit}_{2500}$. \owls\ is the only code including AGN feedback which has baryon fractions similar to codes with SN feedback only. This difference is in stark contrast to the non-radiative simulations, where $\Upsilon_{\rm gas}(R>R^{\rm crit}_{2500})\sim0.8$ with a moderate scatter. 

\par
Given the systematic differences presented here, a natural question to ask is which code+subgrid physics is in reasonable agreement with observations of the cluster environment, especially with the aim of using the gas fractions of simulated clusters for cosmological purposes. 
As pointed out by various studies on the gas fraction of galaxy clusters based on X-ray observations (e.g. \citealt{LaRoque06}; \citealt{Ettori10}; \citealt{Zhang10}; \citealt{Maughan14}), gas is expected to account for around 11-12 per cent of the total mass at $R_{500}^{\rm crit}$, which corresponds to approximately 65-70 per cent of the cosmic ratio.
 
Some AGN codes are in tension with these observations and have $\Upsilon_{\rm gas} >$ 90 per cent (which corresponds to more than 15 per cent of gas with respect to the total mass). 
All the other codes are largely compatible with these results, and these include include methods both with and without AGN feedback.
Moving inward to smaller radii, \cite{Zhang10} and \cite{Vikhlinin09} find lower values (around 9-10 per cent) at $R_{2500}^{\rm crit}$, in keeping with the general trend of falling gas fractions seen in all simulations (whether NR or not).  These values are achived in our comparison by the same set of codes that were found to be in agreement with observational results at $R_{500}^{\rm crit}$.

Nevertheless, in a recent work \cite{Gonzalez13} suggested that massive clusters may have a higher gas content than what was reported by most of observational studies,  estimating a gas fraction around 14 per cent for $M_{500}^{\rm crit}>$ 2$\times$10$^{14}$M$_{\odot}$: these results would support the high gas fraction obtained by codes including AGN, such as \RamsesA, \arepoIL\ and \gx. This is supported also by \cite{Pratt09}, which suggests at the same overdensity $f_{\rm gas} \sim $14 per cent for massive clusters, measuring values up to 16 per cent for individual clusters.

\subsection{Stars}\label{sec:stars}
\begin{figure}
  \includegraphics[width=0.495\textwidth]{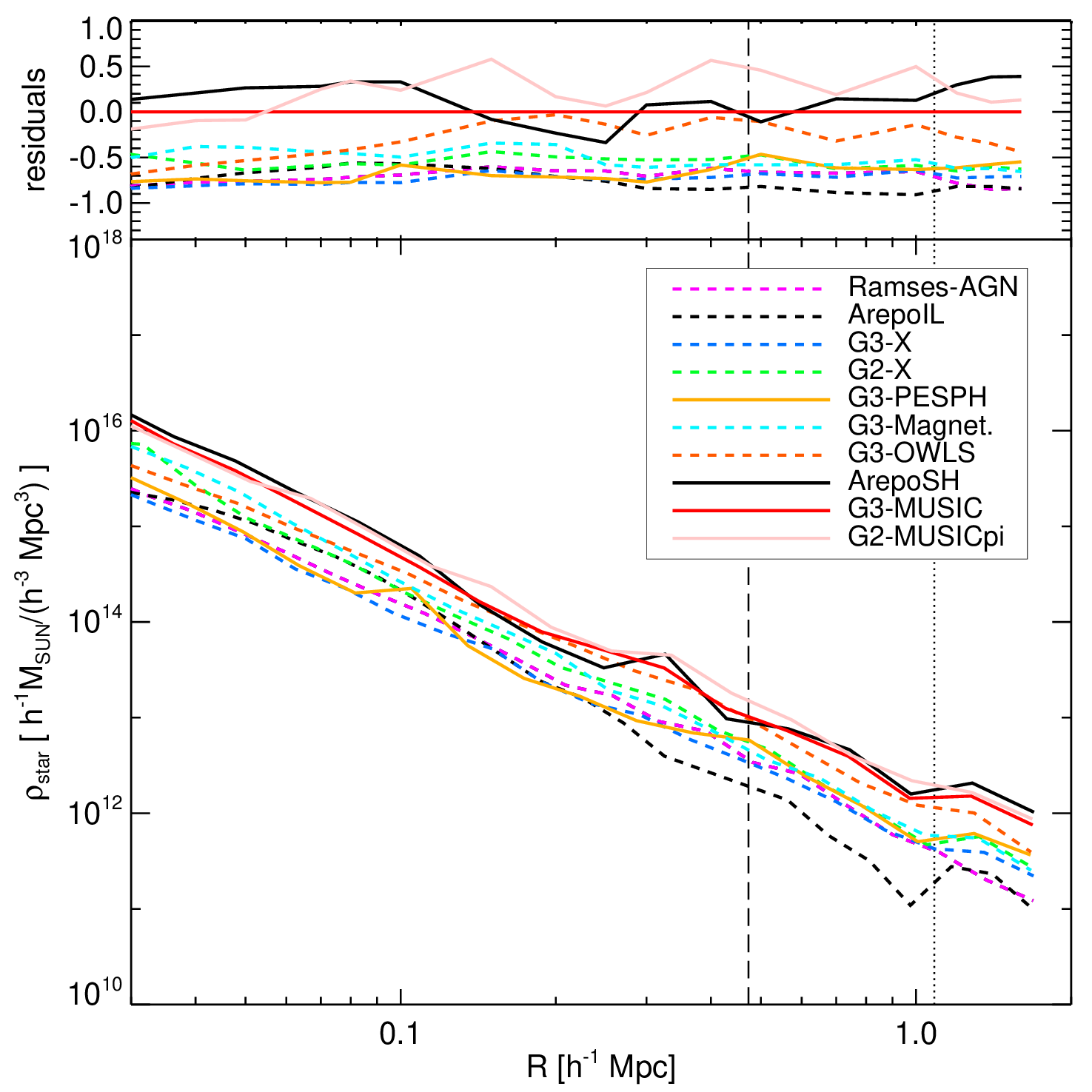}
   \caption{Stellar density profile at $z=0$. Format similar to Fig.~\ref{fig:rho_gas}.}
    \label{fig:rho_star}
\end{figure}

\begin{figure}
  \includegraphics[width=0.476\textwidth]{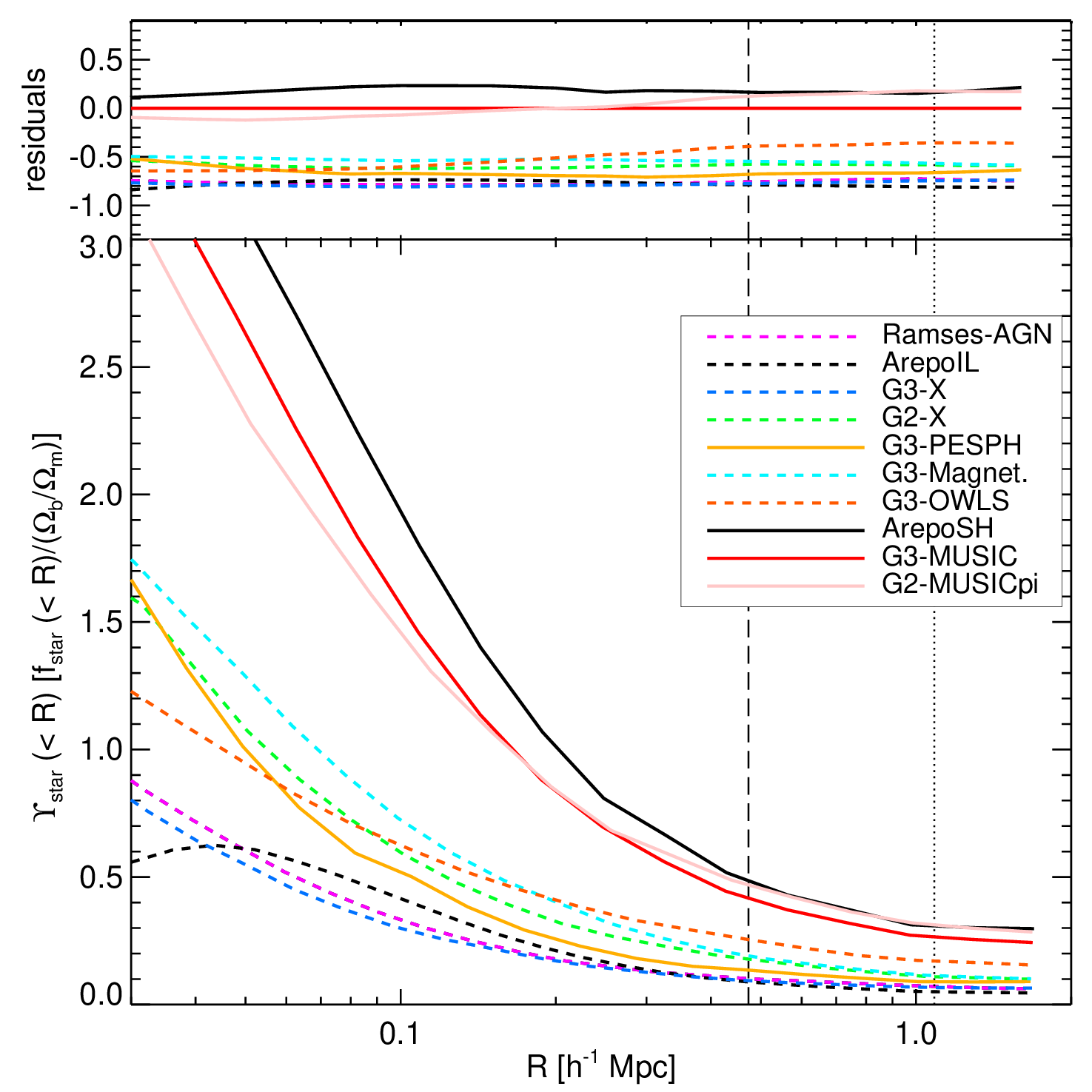}
  \caption{Radial star fraction profile at $z=0$. Format similar to Fig.~\ref{fig:rho_gas}.}
   \label{fig:fstar}
\end{figure}

Here we do not examine the stellar component in detail, i.e., the properties of the galaxies, but defer such an analysis to a companion paper, (Elahi et al., in prep.). Instead we only focus on the overall stellar profiles presented in Figures \ref{fig:rho_star}-\ref{fig:fstar}. These figures show that the stellar distribution does not extend as far as the gas or dark matter distributions and that galaxies dominate the baryonic content of the central regions. As before, however, the profiles show major code-to-code scatter and systematic differences, and generally a clear separation between codes including and $not$ including AGN feedback, with one notable exception, \pesph.
The profiles of the star density are shown in Figure \ref{fig:rho_star}. All the codes which only include stellar feedback and not AGN show very concentrated stellar densities, around a factor of 5 larger than those of the codes which do include AGN. \musicP\ is the code with the highest stellar density within $R^{\rm crit}_{2500}$. 

Unlike the gas densities, the disagreement does not vanish at the cluster outskirts: gas density profiles are mainly determined by gravity in the outskirts, while star formation is determined by local cooling/feedback. The residuals are flat and non-zero out to well past $R^{\rm crit}_{500}$, as shown in the upper panel of Figure \ref{fig:rho_star}; at $R^{\rm crit}_{200}$ there is still an order of magnitude difference between the code with the highest stellar density (\arepoSH) and that with the lowest (\RamsesA).

Similarly to the case of the gas component, we define the star fraction as:
\begin{equation}
    \Upsilon_{\rm star} = \left[\frac{M_{\rm star}(<R)}{M(<R)}\right]\left(\frac{\Omega_b}{\Omega_m}\right)^{-1}.
\end{equation}
and we show the profiles in Figure \ref{fig:fstar}.

Most codes (actually all but \arepoIL, and to a lesser extent \RamsesA\ and \gx) have stellar dominated central regions ($\Upsilon_{\rm star}>$ 1). The importance of AGN feedback in preventing overcooling is indicated by the fact that only codes without AGN feedback typically have $\Upsilon_{\rm star}$ a factor of 2-3 larger than the rest, not only in the cluster core but also in the outskirts. In fact, at $R^{\rm crit}_{2500}$ the codes including AGN feedback already have $\Upsilon_{\rm star} <$ 20 per cent, while for the others the star component still accounts for around 40 per cent of the cosmic ratio. At $R^{\rm crit}_{500}$ all the codes with AGN feedback show values of  $\Upsilon_{\rm star}$ below 10 per cent, while codes with only stellar feedback have a mean value of 30 per cent. \RamsesA, \arepoIL\ and \gx\ are the codes which most efficiently reduce star formation, showing  $\Upsilon_{\rm star} <$ 10 per cent already at $R^{\rm crit}_{2500}$.
Interestingly, \pesph\ is again an outlier in the codes that do not include AGN feedback, having similar $\Upsilon_{\rm star}$ profiles to \gx\ and \gxx . Amongst all the codes, \arepoIL\ is the only one which does not show a monotonically-falling star fraction, exhibiting a small inversion in the cluster core (this may be due to an offset between the BCG and the cluster center). In general, AGN feedback decreases the stellar fraction by a factor of 80-100 per cent. 
\par 
Measurements of the stellar mass of galaxy clusters from observations still do not agree: for massive clusters, \cite{Giodini09} and more recently \cite{Gonzalez13} reported a star fraction between 1 and 2 per cent (corresponding to about 5-10 per cent of the cosmic ratio for WMAP5-WMAP7 cosmologies), while \citealt{Sanderson13} give a value closer to 3 per cent (15-20 per cent of the cosmic baryon fraction). In spite of these discrepancies, observations seem to agree that in massive clusters of galaxies the star mass does not exceed 3 per cent of the total cluster mass. 

Previous works on the baryon contents of hydrodynamical simulations of galaxy clusters have pointed out that methods only including stellar feedback produce an excess of stars (see for instance \citealt{Sembolini2013}) and star fractions not lower than 5 per cent, while codes which take advantage of AGN feedback are able to reproduce stellar masses compatible with observations (e.g. \citealt{Planelles13}). Other detailed comparisons between hydrodynamic simulations and observations can be found in \cite{McCarthy10} and \cite{lebrun14}.
Our results confirm this trend, as all the codes with AGN feedback succeed in recovering stellar fractions below 3 per cent at $R^{\rm crit}_{500}$. Interestingly, \pesph\ is the only code with only stellar feedback which reports a stellar fraction more similar to that of the methods including AGN feedback. This can be explained considering the wind model adopted by \pesph, which strongly suppresses low-mass galaxies using high mass loading. This results in a slower buildup of massive galaxy progenitors at early epochs and less dry merger growth within cluster environments at later epochs (see e.g \citealt{Oppenheimer10,Dave11a,Dave11b}).

Nevertheless, as discussed in the previous section, some codes with AGN feedback show an excess of gas mass in their outskirts which makes their values for the gas fraction incompatible with observational results. Surprisingly, codes that include cooling, star formation, and SN feedback are in better agreement with these observations than some of those that also {\em include AGNs}, which can give baryon fractions that are too high.

It is also interesting that the codes that recover the most realistic results of the gas and stellar fractions are in general those which have been previously calibrated with observations: for instance, in \owls\ the AGN heating temperature has been tuned in order to synthetically reproduce X-ray, SZ and optical cluster properties matching with observations (see \citealt{lebrun14}); \gxx\ calibrated its AGN model to match the pressure profiles measured by Planck \citep{Planck5}, and tuned the SN feedback parameters to get reasonable agreement with the gas and star fractions (see \citealt{Pike14}).


\section{Summary \& conclusions} \label{sec:summary} 
This work is the second paper of the {\it nIFTy cluster comparison project} series. In the first nIFTy paper, 13 different codes have been used to simulate the same massive cluster, describing the baryon component only by means of non-radiative hydrodynamics: we showed that modern SPH codes are able to reproduce the same results as grid-based codes - same gas density and temperature profiles and a large constant entropy core (S15).

Here, we have studied how cluster properties and code-to-code discrepancies change when the the realism of the description of the baryon component is improved by adding radiative mechanisms - such as cooling, star formation, supernovae feedback, black hole accretion and AGN feedback. We have investigated the performance of \ncodes\ modern astrophysical
simulation codes -  \ramses, 2 versions of \arepo\, and \ngadget\ versions of \gadget\ with different SPH implementations. All the simulations have been run using a common set of parameters (e.g. time step accuracy, gravitational softening, dimension of the particle mesh) adopted for S15, but allowing each method to choose the radiative processes modeled by subgrid prescriptions.

We find that - in contrast with what we reported for non-radiative comparison- the differences between classic SPH, modern SPH and grid-based codes are now washed away by the differences in the subgrid physics. The main discrepancies are between codes which include AGN feedback and those which only consider stellar feedback. For instance, the two versions of \arepo\ show significantly different results in the gas and star fraction. Nevertheless, AGN feedback does not always play a dominant role: in particular, entropy profiles do not seem to be sensitive to the inclusion of AGN feedback. Nevertheless, the addition of radiative models seems to drastically change the entropy cores produced by different codes.

\par

Our main results can be summarized as follows:
\begin{itemize}
\item Global properties of the cluster -such as the total mass or shape- as calculated at $R^{\rm crit}_{200}$ are recovered by all the codes with little scatter (less than 2 per cent) and values extremely close to the non-radiative case. The discrepancies are more evident when we consider baryon properties: temperatures are on average 20 per cent higher than the NR runs with a scatter of 5 per cent. The star fraction - which is the global property most strongly dependent on the chosen subgrid physics- has a scatter larger than 60 per cent.
\item Although all the codes (except for \RamsesA\ in some cases) agree well on main the features of the cluster, the dark matter distribution appears to have larger scatter amongst the codes than in the of case of the dark matter-only and non- radiative models. This happens in the central regions,  where the gravitational field of the cluster is dominated by stars. The dark matter is pulled in by cooling baryons and stars. The codes which do not include AGN feedback, which are those with the highest cooling rates and star fraction in the center, are therefore the ones with the highest dark matter concentrations in the core.
\item The gas density profiles show a larger scatter than in the non-radiative case. In this case, we do not observe any difference between grid-based, modern and classic SPH codes; similarly, AGN feedback does not seem to play a dominant role.
\item Temperatures are higher than in the non-radiative case and have a large scatter. Some of the codes show a central temperature inversion, similar to that observed in non-radiative, classic SPH simulations; other codes have a temperature which behaves monotonically, as for modern SPH codes in the non-radiative case. Interestingly, in the full physics case some codes which in their non-radiative version were exhibiting a central inversion of the temperature now show a monotonically decreasing profile (e.g. \music).
\item Entropy profiles are strongly affected by radiative processes and they present a completely different scenario than in the non-radiative case. The differences between classic SPH codes , which showed an entropy profile falling towards the cluster center, and grid-based and modern SPH codes, which showed a flat entropy core, have now disappeared. Most of codes produce an Õalmost-flatÕ central entropy profile, matching the overall X-ray profiles observed. AGN feedback is not necessary to flatten entropy profiles.
\item As expected, codes including AGN are able to limit the problem of overcooling and produce a star content compatible with observations. Codes with only stellar feedback show extremely star-dense cores and an excess of stars also in the outskirts. 

AGN feedback seems also to increase the fraction of gas at all the cluster radii, especially in the outskirts. Comparison with observational studies of the gas fraction at $R^{\rm crit}_{2500}$ and $R^{\rm crit}_{500}$ show that some of the codes including AGN produce an excess of gas of around 15-20 per cent. Codes which have been previously calibrated with observations generally get more realistic results for the gas and star fractions.

\end{itemize}

The next papers of the nIFTy cluster comparison series will investigate in more detail how the different codes and physical mechanisms adopted describe a wide range of cluster properties: an upcoming work (Elahi et al., in preparation) will study more deeply the properties of the star component and of substructures; another work (Cui et al., in preparation) will focus on the differences between dark-matter only, non radiative and full physics runs. Subsequent papers will look at
the recovery of cluster properties such as X-ray temperature and
Sunyaev-ZelÕdovich profiles, gravitational lensing, the cluster outskirts
and hydrostatic-mass bias, all of which will add to our understanding of how consistently the results of different codes can inform our understanding of galaxy cluster properties.

\section*{Acknowledgments}
The authors would like to express special thanks to the Instituto de Fisica Teorica (IFT-UAM/CSIC in Madrid) for its hospitality and support, via the Centro de Excelencia Severo Ochoa Program under Grant No. SEV-2012-0249, during the three week workshop "nIFTy Cosmology" where this work developed. We further acknowledge the financial support of the University of Western 2014 Australia Research Collaboration Award for "Fast Approximate Synthetic Universes for the SKA", the ARC Centre of Excellence for All Sky Astrophysics (CAASTRO) grant number CE110001020, and the two ARC Discovery Projects DP130100117 and DP140100198. We also recognize support from the Universidad Autonoma de Madrid (UAM) for the workshop infrastructure.

GY and FS acknowledge support from MINECO (Spain ) through the grant AYA 2012-31101.
GY  also thanks  the  Red Espa–ola de Supercomputaci—n  for granting the computing time in  the Marenostrum Supercomputer  at BSC. AK is supported by the {\it Ministerio de Econom\'ia y Competitividad} (MINECO) in Spain through grant AYA2012-31101 as well as the Consolider-Ingenio 2010 Programme of the {\it Spanish Ministerio de Ciencia e Innovaci\'on} (MICINN) under grant MultiDark CSD2009-00064. He also acknowledges support from the {\it Australian Research Council} (ARC) grants DP130100117 and DP140100198. He further thanks Radio Dept for lesser matters.. CP acknowledges support of the Australian Research Council (ARC) through Future Fellowship FT130100041 and Discovery Project DP140100198. WC and CP acknowledge support of ARC DP130100117. EP acknowledges support by the Kavli Foundation and the ERC grant ÒThe Emergence of Structure during the epoch of ReionizationÓ. STK acknowledges support from STFC through grant ST/L000768/1. RJT acknowledges support via a Discovery Grant from NSERC and the Canada Research Chairs program. Simulations were run on the CFI-NSRIT funded Saint Mary's Computational Astrophysics Laboratory. SB \& GM acknowledge support from the PRIN-MIUR 2012 Grant "The Evolution of Cosmic Baryons" funded by the Italian Minister of University and Research, by the PRIN-INAF 2012 Grant "Multi-scale Simulations of Cosmic Structures", by the INFN INDARK Grant and by the "Consorzio per la Fisica di Trieste". IGM acknowledges support from a STFC Advanced Fellowship. DN, KN, and EL are supported in part by NSF AST-1009811, NASA ATP NNX11AE07G, NASA Chandra grants GO213004B and TM4-15007X, the Research Corporation, and by the facilities and staff of the Yale University Faculty of Arts and Sciences High Performance Computing Center. PJE is supported by the SSimPL program and the Sydney Institute for Astronomy (SIfA), DP130100117. JIR acknowledges support from SNF grant PP00P2 128540/1. CDV acknowledges financial support from the Spanish Ministry of Economy and Competitiveness (MINECO) through the 2011 Severo Ochoa Program MINECO SEV-2011-0187 and the AYA2013-46886-P grant. AMB is supported by the DFG Research Unit 1254 'Magnetisation of Interstellar and Intergalactic Media' and by the DFG Cluster of Excellence 'Origin and Structure of the Universe'. RDAN acknowledges the support received from the Jim Buckee Fellowship. The AREPO simulations were performed with resources awarded through STFCs DiRAC initiative. The authors thank Volger Springel for helpful discussions and for making AREPO and the original GADGET version available for this project. JS acknowledges support from the European Research Council under the European UnionÕs Seventh Framework Programme (FP7/2007-2013)/ERC grant agreement 278594-GasAroundGalaxies. \pesph\ Simulations were carried out using resources at the Center for High Performance Computing in Cape Town, South Africa

The authors contributed to this paper in the following ways: FS, GY, FRP, AK, CP, STK \& WC formed the core team that organized and analyzed the data, made the plots and wrote the paper. AK, GY \& FRP organized the nIFTy workshop at which this program was initiated. GY supplied the initial conditions. PJE assisted with the analysis. All the other authors, as listed in Section~\ref{sec:codes} performed the simulations using their codes. All authors had the opportunity to proof read and comment on the paper. 

The simulation used for this paper has been run on Marenostrum supercomputer and is publicly available at the MUSIC website\footnote{\href{http://music.ft.uam.es}{http://music.ft.uam.es}}.

This research has made use of NASA's Astrophysics Data System (ADS) and the arXiv preprint server.

\bibliographystyle{mn2e}

\label{lastpage}

\end{document}